\RequirePackage{fix-cm}

\documentclass[smallextended]{svjour3}

\usepackage{natbib}
\usepackage{balance}
\usepackage{fancyhdr}
\usepackage{amssymb}
\usepackage{amsmath}
\usepackage{wrapfig}
\usepackage{multirow}
\usepackage{graphicx}
\usepackage{algorithm}
\usepackage{algorithmic}
\usepackage{url}
\usepackage{booktabs}
\usepackage{color}
\usepackage{moreverb}
\usepackage{balance}
\usepackage{amssymb,amsmath}
\usepackage{wrapfig}
\usepackage{multirow}
\usepackage{graphicx}
\usepackage{algorithm}
\usepackage{algorithmic}
\usepackage{epstopdf}
\usepackage{url}
\usepackage{enumitem}
\usepackage{miniplot}
\usepackage{ifthen}

\usepackage{pifont}
\usepackage{bbding}
\usepackage{booktabs}
\usepackage{threeparttable}
\usepackage{array}
\usepackage{ragged2e}
\usepackage{etoolbox}
\makeatletter

\patchcmd{\NAT@citex}
  {\@citea\NAT@hyper@{%
     \NAT@nmfmt{\NAT@nm}%
     \hyper@natlinkbreak{\NAT@aysep\NAT@spacechar}{\@citeb\@extra@b@citeb}%
     \NAT@date}}
  {\@citea\NAT@nmfmt{\NAT@nm}%
   \NAT@aysep\NAT@spacechar\NAT@hyper@{\NAT@date}}{}{}

\patchcmd{\NAT@citex}
  {\@citea\NAT@hyper@{%
     \NAT@nmfmt{\NAT@nm}%
     \hyper@natlinkbreak{\NAT@spacechar\NAT@@open\if*#1*\else#1\NAT@spacechar\fi}%
       {\@citeb\@extra@b@citeb}%
     \NAT@date}}
  {\@citea\NAT@nmfmt{\NAT@nm}%
   \NAT@spacechar\NAT@@open\if*#1*\else#1\NAT@spacechar\fi\NAT@hyper@{\NAT@date}}
  {}{}

\makeatother

\newboolean{showcomments}
\setboolean{showcomments}{true}
\ifthenelse{\boolean{showcomments}}
{ \newcommand{\mynote}[2]{
    \fbox{\bfseries\sffamily\scriptsize#1}
    {\small$\blacktriangleright$\textsf{\emph{#2}}$\blacktriangleleft$}}}
{ \newcommand{\mynote}[2]{}}

\ifthenelse{\boolean{showcomments}}
{
  
}{
  \newcommand{\mynote}[2]{}
}

\definecolor{orange}{rgb}{1,0.5,0}

\begin{document}

\title{What Makes a Popular Academic AI Repository?}
\titlerunning{What Makes a Popular Academic AI Repository?}

  \author{Yuanrui Fan   \and Xin Xia \and David Lo \and
        Ahmed E. Hassan \and Shanping Li}

\institute{Yuanrui Fan \at
         College of Computer Science and Technology, Zhejiang University, China\\
         PengCheng Laboratory, China\\
         \email{yrfan@zju.edu.cn}
         \and
         Xin Xia \at
         Faculty of Information Technology, Monash University, Australia \\
         \email{xin.xia@monash.edu}
         \and
         David Lo \at
         School of Information Systems, Singapore Management University, Singapore\\
         \email{davidlo@smu.edu.sg}
         \and
         Ahmed E. Hassan \at
         School of Computing, Queen's University, Canada\\
         \email{ahmed@cs.queensu.ca}
         \and
         Shanping Li \at
         College of Computer Science and Technology, Zhejiang University, China\\
         \email{shan@zju.edu.cn}}

\date{Received: date / Accepted: date}
\maketitle
\maketitle
\thispagestyle{fancy}
\lhead{}
\chead{}
\rhead{}
\lfoot{}
\cfoot{}
\cfoot{\thepage}
\renewcommand{\headrulewidth}{0pt}
\renewcommand{\footrulewidth}{0pt}
\pagestyle{fancy}
\cfoot{\thepage}

\begin{abstract}
  Many AI researchers are publishing code, data and other resources that accompany their papers in GitHub repositories. In this paper, we refer to these repositories as academic AI repositories. Our preliminary study shows that highly cited papers are more likely to have popular academic AI repositories (and vice versa). Hence, in this study, we perform an empirical study on academic AI repositories to highlight good \emph{software engineering practices} of popular academic AI repositories for AI researchers.

We collect 1,149 academic AI repositories, in which we label the top 20\% repositories that have the most number of stars as popular, and we label the bottom 70\% repositories as unpopular. The remaining 10\% repositories are set as a gap between popular and unpopular academic AI repositories. We propose 21 features to characterize the software engineering practices of academic AI repositories. Our experimental results show that popular and unpopular academic AI repositories are statistically significantly different in 11 of the studied features---indicating that the two groups of repositories have significantly different software engineering practices. Furthermore, we find that the number of links to other GitHub repositories in the README file, the number of images in the README file and the inclusion of a license are the most important features for differentiating the two groups of academic AI repositories. Our dataset and code are made publicly available to share with the community.

\end{abstract}


\section{Introduction}
Transparency, openness, and reproducibility are considered as crucial features of science~\citep{nosek2015promoting}. In the area of Artificial Intelligence (AI), researchers are encouraged to publish data, software code and other resources that accompany their papers~\citep{sonnenburg2007need}. Prior AI studies have commonly used GitHub repositories to share such resources~\citep{NIPS2018_7391,yang2018graph}. We refer to GitHub repositories that accompany AI publications as \emph{academic AI repositories}.

As a social coding platform, GitHub provides the \emph{starring} feature for users to express their interest or satisfaction with a repository. Number of stars is commonly used as a proxy for measuring the popularity of a repository~\citep{borges2016predicting,borges2016understanding, hu2016influence, han2019characterization}. Following these studies, we use the number of stars to quantify the popularity of an academic AI repository. We hypothesize that popularity of an academic AI repository is an indicator that the corresponding research attracts great interest. In this study, we set out to investigate characteristics of popular academic AI repositories with respect to their software engineering practices. Notice that our aim is \emph{not} to uncover causal relationships between various factors and popularity of academic AI repositories. Instead, we compare the differences between popular and unpopular academic AI repositories, aiming to help highlight good software engineering practices in publishing AI code or data for the community, e.g., documentation, choice of programming language, etc.

In this study, we collected 1,149 academic AI repositories. Among the repositories, we observe that many academic AI repositories are not popular. For instance, 178 of the repositories receive fewer than 10 stars. Thus, many AI researchers may thrive to understand how to make their academic AI repositories receive more attention. Furthermore, based on our collected data, we find that the number of stars of the academic AI repositories and the number of Google Scholar citations of the corresponding papers show a high correlation with a Spearman's $\rho$~\citep{zar2005spearman} of above 0.7---indicating that highly cited papers are more likely to have popular academic AI repositories (and vice versa). Although association does not imply causation, it is still worth highlighting useful software engineering practices of managing academic AI repositories from high-impact AI research.

We propose 21 features to characterize the software engineering practices of academic AI repositories. These features are grouped along three dimensions: code, reproducibility, and documentation. The three dimensions are derived from prior popularity analyses of GitHub repositories (e.g.,~\cite{borges2016understanding}) and our preliminary analysis of the academic AI repositories in our dataset. Notice that we focus our analysis on the features that \emph{AI researchers can control and strive to improve or change in their repositories} following~\citet{tian2015characteristics}. Hence, features like number of contributors and repository owner (e.g., organization or personal) are not considered in this study because these features cannot be easily changed by repository owners.

To analyze the difference between popular and unpopular academic AI repositories, we need to label the studied repositories as popular or unpopular according to the number of stars. We notice that the distribution of number of stars in our repository dataset is highly skewed. We label the top 20\% repositories that have the most number of stars as popular following prior studies~\citep{phua2004minority,alves2010deriving,yan2017automating}. And we label the bottom 70\% repositories as unpopular. The remaining 10\% repositories are set as a gap between popular and unpopular academic AI repositories. The top 20\% repositories have at least 154 stars, and the bottom 70\% have at most 99 stars. With the gap, the difference concerning the number of stars between the two groups of repositories is at least 55 stars.

Through an empirical analysis on the labeled academic AI repositories, we answer the following two research questions:

\vspace{0.1cm}\noindent\textbf{RQ1: Is there a relationship between each feature and the popularity of academic AI repositories?}

For each feature, we leverage Wilcoxon rank-sum test and Cliff's delta to investigate whether popular academic AI repositories are different from unpopular ones with respect to the feature. We find that popular academic AI repositories are statistically significantly different from unpopular ones with a non-negligible effect size in 11 out of our 21 features. Hence, the two groups of repositories have significantly different software engineering practices. Generally, popular academic AI repositories contain more code files, code lines, and modules. They take more measures to ease the reproducibility of results reported in the corresponding papers, e.g., the inclusion of pre-trained models and running scripts. Their documentation is better organized (e.g., by using more lists) and provides more detailed information about the repository (e.g., by presenting more images).

\vspace{0.1cm}\noindent\textbf{RQ2: What are the most important features that differentiate popular and unpopular academic AI repositories?}

To investigate the most important features, we learn a random forest classifier and use the extracted features as input features. We remove correlated and redundant features to avoid negative impact of these features on interpreting models. Our analysis shows that the number of links to other GitHub repositories in the README file, the number of images in the README file and inclusion of a license are the most important features that can differentiate the two groups of academic AI repositories.

Our main contributions can be summarized as follows.

\begin{itemize}
\item To our knowledge, we are the first to study the popularity of academic AI repositories. Through an empirical analysis of more than 1,000 academic AI repositories, we highlight good software engineering practices of popular academic AI repositories that can help AI researchers improve their own academic AI repositories. We find that popular academic AI repositories take more measures to ease the reproducibility of results that are reported in the corresponding paper. Their documentation is better organized and provides more detailed information about the repository.  Furthermore, we find that popular academic AI repositories contain more code files, code lines and modules.

\item Furthermore, the dataset and code that accompany our study are now available on GitHub to share with and benefit the community.\footnote{https://github.com/YuanruiZJU/academic-ai-repos}

\end{itemize}

\vspace{0.1cm}\noindent\textbf{Paper Organization.} The remainder of this paper is organized as follows. Section~\ref{related} briefly reviews the related work. Section~\ref{data} describes the collection of the studied academic AI repositories and our preliminary analysis of the repositories. Section~\ref{features} elaborate the details of the studied repository features. Section~\ref{result} presents the results with respect to the two research questions. Section~\ref{discuss} discloses the threats to validity of our study. Finally, Section~\ref{conclusion} concludes our paper.

\section{Background and Related Work}\label{related}
In this section, we first give a brief overview of open source in the area of AI. Then, we highlight the studies that are related to our work.

\vspace{0.1cm}\noindent\textbf{Open Source in the Area of AI.} Open source has been advocated in the area of AI for many years.~\citet{sonnenburg2007need} noted the necessity of open sourcing machine learning projects.~\cite{collberg2016repeatability} observed that publications may not describe the details of used variables, parameters or functions. They found that computer systems research is difficult to reproduce when prototype systems are unavailable.~\cite{gundersen2017reproducible} highlighted ten benefits of making reproducible AI research for AI researchers, e.g., receive credit for research products and increase the number of citations.

Nowadays, many AI researchers publish their code, data and other resources that accompany their papers as GitHub repositories~\citep{NIPS2018_7391,yang2018graph}. These repositories are referred to as academic AI repositories in this study. Many AI studies based their implementations on code in academic AI repositories. For example,~\citet{yang2018graph}'s work is based on seven academic AI repositories (see their accompanying academic AI repository\footnote{https://github.com/jwyang/graph-rcnn.pytorch}). Academic AI repositories are playing an important role in the community.

The popularity of an academic AI repository may indicate the attractiveness of the repository and the corresponding publication for other researchers and users. In this study, we investigate the characteristics of popular academic AI repositories, aiming to highlight good practices of publishing academic repositories for AI researchers.

\vspace{0.1cm}\noindent\textbf{Studies on the Popularity of GitHub Repositories.} Many studies have analyzed software code repositories to understand what makes a code repository popular.~\citet{weber2014makes} attempted to differentiate popular and unpopular GitHub repositories written in Python. They showed that the features characterizing repositories outperform those characterizing repository authors.~\citet{zhu2014patterns} analyzed the folder use of 140k GitHub repositories, and they found that the use of standard folders (e.g., doc) may increase the likelihood of a repository being popular.~\citet{aggarwal2014co} observed that popular repositories attract more documentation collaborators.~\citet{bissyande2013popularity} collected a dataset of 100K open source projects and they analyzed popularity, interoperability and impact of various programming languages.~\citet{borges2016understanding} analyzed 2,500 repositories that have the most number of stars in GitHub and observed that application domain can impact the popularity of GitHub repositories.~\citet{han2019characterization} collected more than 400K GitHub repositories, and they extracted properties of the repositories such as number of branches and number of contributors. Then they studied the relationship between the repository properties and the popularity of the repositories.

Different from the above studies, we focus our analysis on the popularity of academic AI repositories. To our best knowledge, we are the first to study academic AI repositories. We also analyze the relationship of additional features that are special to academic AI repositories with the popularity of these repositories, e.g., features that characterize reproducibility of the corresponding paper. In comparison with the above studies, we focus on the features that AI researchers can control or strive to improve in their repositories, aiming to give actionable suggestions for academic AI repositories. For example, although~\cite{aggarwal2014co} observed that popular projects require more documentation efforts, they did not give specific suggestions. In this study, we propose 10 features to characterize documentation of popular and unpopular academic AI repositories. Based on the studied features, we are able to give more actionable suggestions for AI researchers to improve the documentation of their academic AI repositories.

\section{Data Preparation and Preliminary Study}\label{data}
In this section, we first describe our approach to collecting and cleaning the studied data. Then, we present our preliminary study, in which we investigate the relationship between Google Scholar citations of the AI publications and the popularity of their accompanying repositories.

\subsection{Data Preparation}\label{sec:prepare}

Our study starts from a GitHub repository \texttt{pwc} (i.e., papers with code).\footnote{https://github.com/zziz/pwc} At the time of crawling the repository (April 2019), the repository maintained a list of 1,448 AI papers that were published in 2015--2018 with references to their corresponding GitHub repositories. Each record in the list includes the following pieces of information: paper title, conference name, publication and link to the corresponding repository. The listed papers were published in NeurIPS, ICML, CVPR, ICCV and ECCV. Based on the listed papers and repositories, we take three steps to prepare the studied data:

\vspace{0.1cm}\noindent\textbf{Data Cleanining:} We observed that not all of the listed repositories were created by the original paper authors. These repositories are third-party implementations of the papers or not related with the papers. For example, the linked code repository of the paper ``Learning to Branch'' is the \texttt{github-workflow-activity} repository,\footnote{https://github.com/foss2serve/github-workflow-activity} which is a repository to help users learn the GitHub workflow activities. However, the paper ``Learning to Branch'' is related to tree search and tree partitioning techniques~\citep{balcan2018learning}. To retrieve a clean dataset, we manually select academic AI repositories from those listed by the \texttt{pwc} repository. Given one paper and its corresponding repository, we take the following two steps to make our decision:

\vspace{0.05cm}\underline{\em Step 1: Check README file.} We first check whether the README file of the repository contains words like ``our work'' and ``our paper''. The appearance of such words indicates that the paper authors are contributing the repository. Hence, we consider the repository as an academic AI repository. If the README file does not contain such words, we take the next step.

\vspace{0.05cm}\underline{\em Step 2: Check Contributor Usernames.} We retrieve names of the paper authors and contributor usernames of the repository. Then, we compare each of the paper authors' name and each of contributors' username. If one of the contributors' username is similar to a paper author name (e.g., the contributor username is the abbreviation of the paper author name), we consider that the contributor is the paper author, and the repository is an academic AI repository.

The first two authors of this paper manually checked these papers and repositories independently. And each author spent about 16 hours in labelling whether each of the repositories is an academic AI repository. We apply Fleiss Kappa~\citep{fleiss1971measuring} to estimate the agreement between the two annotators. The Kappa value is 0.97---indicating that the two annotators have almost perfect agreement. For a few papers and corresponding repositories, the two authors disagree with each other. For example, for the paper ``Cost efficient gradient boosting'', the corresponding repository is the \texttt{LightGBM-CEGB} repository,\footnote{https://github.com/svenpeter42/LightGBM-CEGB} which is a fork of the \texttt{LightGBM} repository.\footnote{https://github.com/microsoft/LightGBM} One annotator considers that the \texttt{LightGBM-CEGB} repository is not an academic repository  because the contributors of the repository do not even change the README file (i.e., the README files of the two repositories are identical). For such repositories, the two authors discuss to determine whether to consider them as academic AI repositories. Among the 1,448 repositories that are listed by the \texttt{pwc} repository, we find that 299 ones are not contributed by the original paper authors. We exclude them from our study. Finally, we get a list of 1,149 academic AI repositories with their related papers, including 211 NeurIPS papers, 116 ICML papers, 527 CVPR papers, 160 ICCV papers and 135 ECCV papers which were published in 2015--2018.

\vspace{0.1cm}\noindent\textbf{Data Crawling:} For each of the 1,149 academic AI repositories, we leverage the \texttt{git clone} command to download the repository data to local disk. Moreover, for all of the studied repositories, we retrieve their meta-data information including license and number of stars via the GitHub REST API.\footnote{https://developer.github.com/v3/} The data is collected on April 2019.

\begin{figure}
  \centering
   \includegraphics[width=6.5cm]{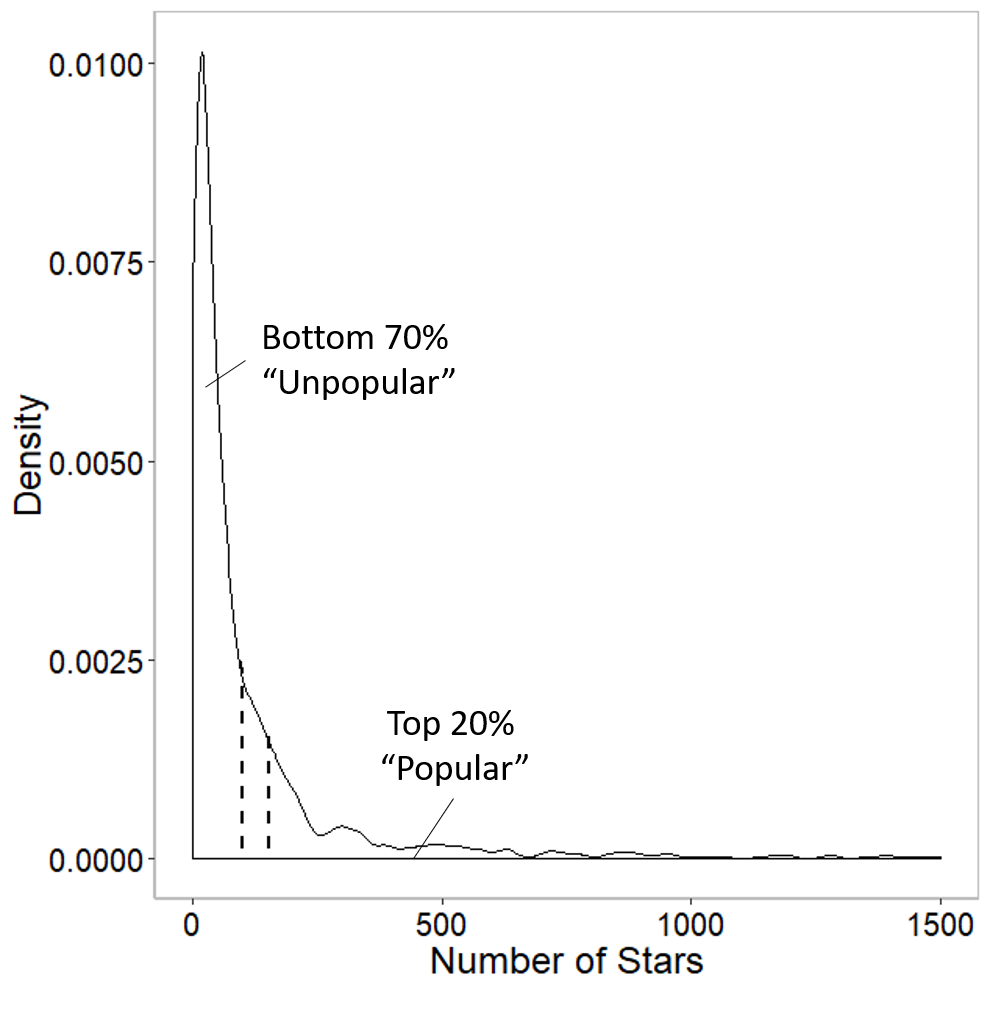}\\
  \caption{Distribution of the number of stars for our studied academic AI repository dataset. Note that for better visibility, repositories that receive more than 1,500 stars are not shown in the figure.}\label{density}
\end{figure}

\vspace{0.1cm}\noindent\textbf{Data Labelling:} To conduct our analysis, we need to categorize repositories as {\em popular} or {\em unpopular} based on their number of stars. Figure~\ref{density} plots the density distribution of the number of stars for our academic AI repository dataset. We find that the number of stars is highly skewed and only a small proportion of repositories received many stars. The 20/80 boundary is commonly used to split skewed data in prior studies~\citep{phua2004minority,alves2010deriving,yan2017automating}. Following these studies, we label top 20\% repositories that have the most number of stars as \emph{popular}. We label the bottom 70\% repositories as \emph{unpopular}. The remaining 10\% repositories are used as a gap between popular and unpopular academic AI repositories. Using the gap, we make the difference with respect to the number of stars between the two groups of repositories more considerable. The top 20\% repositories have at least 154 stars and the bottom 70\% repositories have at most 99 stars, i.e., the difference with respect to the number of stars between the two groups of repositories is at least 55 stars. In total, 1,040 academic AI repositories are labeled.

\subsection{Preliminary Study}\label{sec:preliminary}

For papers with accompanying academic AI repositories, two measures can be used to evaluate the impact or popularity of the research work described in the paper, i.e., the citation statistics (e.g., Google Scholar citation) of the paper and the number of stars of the academic AI repository. Researchers may care more about the citation statistics. An interesting question about such papers and repositories is that: ``is there any relationship between the paper citation and the number of stars of the academic AI repositories?''.

To answer the above question, we collect the Google Scholar citations of papers in our dataset. Notice that the citation data were also collected on April 2019. Papers published longer are likely to get more citations. Hence, we group the papers in our dataset according to their publishing year. For papers published in each year, we calculate the correlation between their citations and the number of stars of their accompanying academic repositories using the Spearman rank correlation test~\citep{zar2005spearman}. We drop the papers published in 2018 since they may not get enough citations.

Table~\ref{spearman} presents the results of the Spearman rank correlation tests. The statistical tests show that for papers published in each year, their citations and number of stars of the papers' accompanying repositories are strongly correlated (with Spearman's $\rho>0.7$ and p-value $<0.001$). Hence, highly cited papers are likely to have popular academic repositories (and vice versa). Although association does not imply causation, for researchers who thrive to improve the impact of their research, this finding still encourages them to learn software engineering practices of popular academic AI repositories and improve their own academic AI repositories.

\begin{table}
  \centering
  \caption{Correlation between citations of papers and number of stars of the papers' accompanying repositories}\label{spearman}
  \begin{tabular}{|c|c|c|c|}
  \hline
  \textbf{Year} & \textbf{\#Papers} & \textbf{Spearman's $\rho$} & \textbf{p-value} \\
  \hline
  2017 & 503 & 0.72 & $<$0.001 \\
  2016 & 70 & 0.70 & $<$0.001 \\
  2015 & 62 & 0.73 & $<$0.001 \\
  \hline
  \end{tabular}
\end{table}

\section{Repository Features}\label{features}

\begin{table*}
	\centering
	\caption{Studied repository features}\label{tab:feature}
    \scriptsize
	\begin{tabular}{|c|l|l|p{1.5in}|p{2in}|}
        \hline
        \textbf{Dimension} & \textbf{Feature Name} & \textbf{Type} & \textbf{Description} & \textbf{Rationale}\tabularnewline
        \hline
        \multirow{18}{*}{\textbf{Code}} & num-code-file &  Numeric & Number of code files & A small number of code files might indicate the code of the repository is not detailed enough.\\
            \cline{2-5}
            & num-code-line & Numeric & Number of lines of code & A small number of code lines might indicate the code of the repository is not detailed enough.\\
            \cline{2-5}
             & num-module &  Numeric & Number of sub-directories in the root directory & \multirow{4}{2in}{Software modularization plays a significant part in ease of code comprehension~\citep{woodfield1981effect}.}\\
             \cline{2-4}
             & num-root-file & Numeric & Number of files in the root directory & \\
             \cline{2-5}
             & language & String & Mainly used programming language & Programming language may impact popularity of GitHub repositories~\citep{borges2016understanding}. \\
             \cline{2-5}
             & framework & String & Machine learning framework & Repositories using popular ML frameworks may attract more attention.\\
             \cline{2-5}
             & num-copy-paste & Numeric & Number of copy-paste code blocks & Bugs can be propagated by copying and pasting code~\citep{kim2004ethnographic,li2006cp}. \\
        \hline
        \multirow{8}{*}{\textbf{Reproducibility}} & has-data & Boolean & Whether the repository includes a dataset & \multirow{8}{2in}{Academic AI repositories make the reproducibility of scientific publications easier~\citep{sonnenburg2007need}. Repositories that can easily reproduce paper results are likely to attract more attention.}\\
                        \cline{2-4}
                        & has-docker & Boolean & Whether the repository uses Docker & \\
                        \cline{2-4}
                        & has-model & Boolean & Whether the repository contains pre-trained models& \\
                        \cline{2-4}
                        & has-shell & Boolean & Whether the repository contains shell scripts & \\
        \hline
        \multirow{21}{*}{\textbf{Documentation}} & num-list & Numeric & Number of lists in the README file & Using list is recommended for clear writing~\citep{kimble1992plain}.\\
                      \cline{2-5}
                      & num-code-blk & Numeric & Number of code blocks in the README file & Users can operate easily by copying and pasting code blocks.\\
                      \cline{2-5}
                      & num-inline-code & Numeric & Number of inline code elements in the README file & A README file with more inline code elements may introduce more details about the repository.\\
                      \cline{2-5}
                      & num-img & Numeric & Number of images in the README file & \multirow{8}{2in}{Showing more results that are produced by the code might attract more attention.}\\
                      \cline{2-4}
                      & num-ani-img & Numeric & Number of animated images in the README file & \\
                      \cline{2-4}
                      & has-video & Boolean & Whether the README file contains link to a video & \\
                      \cline{2-4}
                      & num-table & Numeric & Number of tables in the README file & \\
                      \cline{2-5}
                      & num-ghlink & Numeric & Number of GitHub links in the README file & \multirow{5}{2in}{The README file should provide links to further documentation and support channels~\citep{prana2018categorizing}.}\\
                      \cline{2-4}
                      & has-project-page & Boolean & Whether the README file contains a URL to a project page & \\
                      \cline{2-5}
                      & has-license & Boolean & Whether the repository has a license & Users can clearly know how to use a repository when the repository has a license~\citep{sonnenburg2007need}.\\
        \hline

	\end{tabular}
\end{table*}

In this section, we elaborate on the studied repository features. We extract 21 features that may be potentially associated with the popularity of an academic AI repository. Table~\ref{tab:feature} summarizes the 21 features, which are grouped along three dimensions: code, reproducibility, and documentation. The three dimensions are derived based on prior work on the popularity of general-purpose GitHub repositories in literature (e.g.,~\cite{borges2016understanding}) and our preliminary analysis of the academic AI repositories in our dataset.

We use the 21 features to characterize the software engineering practices of academic AI repositories. For each feature, we compare popular and unpopular academic repositories with respect to the value of the feature. By doing this, we investigate whether popular and unpopular academic AI repositories have different software engineering practices. Furthermore, we use the 21 features to represent each popular and unpopular academic AI repository. Based on the features, we learn a random forest classifier and calculate the most important features for the classifier in differentiating the two groups of repositories.

Notice that many of our features are extracted based on the README file of an academic AI repository. README files are written in Markdown format. To extract features from a README file, we first convert it into HTML format using the Python package \textbf{mistune}.\footnote{https://mistune.readthedocs.io/en/latest/} Next, we extract the structural information from each README file (e.g., images, URLs, lists) by parsing the converted HTML file. For example, we extract any URLs by finding the ``$\langle$a$\rangle$'' tag in the HTML file. For a README file, we denote the set of URLs and the set of sentences containing URLs as $S_{url}$ and $S_{url\_st}$, respectively. In this study, we focus on analyzing the README file in the root directory of studied repositories. The README file in the root directory is the first document that a visitor will see when he/she encounters a new repository~\citep{prana2018categorizing}. It plays an essential role in shaping the visitor's first impression of the repository, which is crucial for the visitor when further navigating the repository~\citep{fogel2005producing}.

\subsection{Code Dimension}

Code dimension refers to features that are based on the code of the repository. In an academic AI repository, code plays an important role.~\citet{sonnenburg2007need} noted that people can use shared code to learn and understand algorithm details that are hidden in publications. \textbf{We expect that academic AI repositories presenting code that is as detailed as possible and easy-to-understand are more likely to be popular.} We use 7 features as listed in Table~\ref{tab:feature} to characterize the code of academic AI repositories.

Publications may not provide details of used variables, parameters, or functions~\citep{collberg2016repeatability}. We expect that an academic AI repository should provide such details that are very useful for others. We use the number of code lines and code files (\emph{num-code-line} and \emph{num-code-file}) to characterize how detailed the presented code by an academic AI repository is. We detect the format of a file by checking its extension. Then we consider the files in the format of one programming language such as C/C++, Python, Java, etc. as code files. Other files such as image and HTML files are dropped since these files are not likely to be used for implementing algorithms. The \emph{num-code-file} feature is calculated by counting the number of code files, and the \emph{num-code-line} feature is calculated by counting the number of lines of code in these files.

\citet{woodfield1981effect} concluded that modularization plays a significant part in ease of code comprehension. We use the number of modules (\emph{num-module}) to characterize the modularization strength of a repository. For a repository, we consider sub-directories in the root directory as modules. We expect that academic AI repositories with more modules (i.e., a larger \emph{num-module}) are better modularized. On the other hand, showing many files in the root directory indicates that files of the repository are not well arranged into different modules. Hence, we also use the number of files in the root directory (\emph{num-root-file}) to quantify the modularization of a repository. 

\citet{borges2016understanding} found that programming language impacts repository popularity. Academic code using popular programming languages can be easily understood by more people who are familiar with the languages. We define the language of an academic AI repository as the mainly used programming language with the most number of files in the repository. To calculate the \emph{language} feature of a repository, we first identify the programming language of each file by checking the file extension. Then, the programming language used by the most number of files is identified as the language of the repository.

Similarly to the \emph{language} feature, academic AI code using popular machine learning frameworks can also be easily understood by more people with knowledge of the frameworks. The used machine learning framework is one of the system requirements of an academic AI repository. In prior studies, README files were used as the source for extracting system requirements~\citep{portugal2016extracting, prana2018categorizing}. Thus, we calculate the \emph{framework} of an academic AI repository from its README file. We first define several candidate frameworks by manually checking the ML frameworks used in our academic AI repository dataset. The candidate frameworks include Caffe, Caffe2, Chainer, MATLAB, MXNet, PyTorch, Tensorflow, and Torch. Then, we extract the used framework from the textual content of the README file (e.g., from ``Requirements'' section of the README file which describes the required environment for running the code). If the README file does not explicitly present any of the candidate frameworks, we set \emph{framework} as \texttt{Unknown}. A README file without explicitly stating the used framework may indicate that the file has low quality. Hence, the \emph{framework} feature can also characterize the documentation quality of an academic AI repository.

Prior studies noted that bugs could be easily propagated by copying and pasting code~\citep{kim2004ethnographic,li2006cp}. Hence, copy-paste code may negatively impact software quality. Low-quality code may discourage other researchers from incorporating the code into their research. We expect that the academic AI repositories with too many copy-paste code blocks are less likely to be popular. Note that we define a code block as code with at least 100 tokens. To calculate the number of copy-paste code blocks (\emph{num-copy-paste}) in an academic AI repository, we apply the copy-paste code detection tool PMD's CPD\footnote{https://pmd.github.io/latest/pmd\_userdocs\_cpd.HTML} on the code of the repository. The \emph{num-copy-paste} feature is calculated as the number of copy-paste code blocks as detected by the tool.

\subsection{Reproducibility Dimension}

Reproducibility dimension refers to the features that measure the difficulty of reproducing the paper results using the code repository. Ease of reproducibility is listed as the first advantage of open sourcing machine learning code~\citep{sonnenburg2007need}. \textbf{We expect that academic AI repositories that take measures to ease the reproducibility of the paper results are more likely to be popular.} We use four binary features as listed in Table~\ref{tab:feature} to characterize the ease of reproducibility of our academic AI repositories. By default, these features are set as 0.

The \emph{has-data} feature represents whether an academic AI repository includes datasets for reproducing results of the paper. The absence of datasets can make reproducing results of the paper difficult since looking for the datasets used in the paper may require much effort. To calculate \emph{has-data}, we try to find a directory in the repository, with its name containing ``data''. If such a directory exists, we set \emph{has-data} to 1. In addition, researchers may provide URLs for downloading datasets in README file. Hence, if a sentence in $S_{url\_st}$ of the README file contains ``data'' or ``dataset'', we set \emph{has-data} to 1.

The \emph{has-docker} feature refers to whether an academic AI repository uses Docker. Docker is a lightweight virtualization software, which can be used to run software packages~\citep{wan2017mining}. Users can bundle applications, tools, libraries and configuration files into a Docker image.~\citet{boettiger2015introduction} recommended the use of Docker for reproducible research. With Docker, academic AI repositories can provide users a convenient way to deploy a running environment. To calculate \emph{has-docker}, we check names of files, and if a file's name contains ``docker'', we set \emph{has-docker} to 1. Also, researchers may provide URLs for downloading their Docker images in README file. Hence, if a sentence in $S_{url\_st}$ of the README file contains ``docker'', we set \emph{has-docker} to 1.

The \emph{has-model} feature represents whether an academic AI repository provides pre-trained models. Building models can be a time-consuming and resource-intensive task. With pre-trained models, users can sidestep troubles of building models when reproducing paper results. We observe that academic AI repositories typically do not provide pre-trained models as a file in the repository since a pre-trained model can be very large. Instead, if AI researchers provide pre-trained models, they mention the models in the README file and provide methods for retrieving the models. For example, the \texttt{vid2vid} repository\footnote{https://github.com/NVIDIA/vid2vid} mentions the provided pre-trained models in the README file and provides commands for downloading the models. Hence, if the README file contains words like ``pre-trained model'' or ``trained model'', we set \emph{has-model} to 1.

The \emph{has-shell} feature represents whether an academic AI repository contains shell scripts. With shell scripts, users can avoid the trouble of inputting commands when installing the environment or running experiments. If the repository contains a file with extension ``.sh'', ``.bash'' or ``.bat'', we set \emph{has-shell} to 1.

\subsection{Documentation Dimension}

Documentation dimension refers to features for characterizing the quality of the documentation (e.g., the README file) in the repository.~\citet{aggarwal2014co} found that popular repositories have well-maintained documentation. \textbf{We expect that repositories with high-quality documentation are more likely to be popular.} We use ten features as listed in Table~\ref{tab:feature} to characterize the documentation quality of an academic AI repository. Most of these features are extracted from README files of academic AI repositories. By default, all of the ten features are set to 0.

The \emph{num-list} feature represents the number of lists in the README file of an academic AI repository.~\citet{kimble1992plain} recommended to use lists for clear writing. Hence, we consider that README files with more lists can more clearly present information. To calculate \emph{num-list}, we count the number of ``$\langle$ol$\rangle$'' and ``$\langle$ul$\rangle$'' tags in the HTML file converted from the README file.

\begin{figure}
  \centering
   \includegraphics[width=12cm]{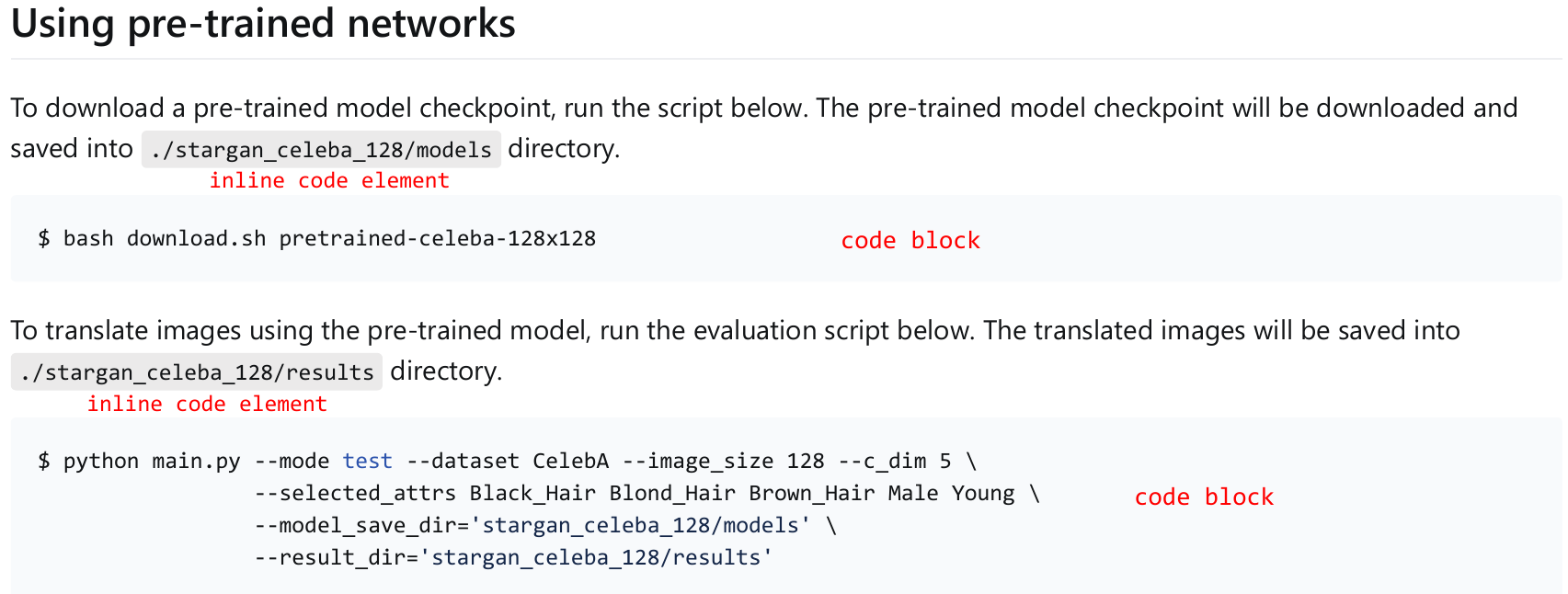}\\
  \caption{A partial screenshot of the README file from the \texttt{StarGAN} repository for showing examples of code blocks and inline code elements.}\label{snapshot}
\end{figure}

The features \emph{num-code-blk} and \emph{num-inline-code} refer to the number of code blocks and inline code elements in the README file of an academic AI repository, respectively. In Figure~\ref{snapshot}, we present a partial screenshot of the README file from the \texttt{StarGAN} repository,\footnote{https://github.com/yunjey/stargan} which is a popular academic AI repository that receives thousands of stars. In the figure, we show examples of code blocks and inline code elements. In a Markdown file, a code block is fenced by lines with three back ticks, while an inline code element is surrounded by two back ticks in a natural language sentence. Code blocks are typically used to present the commands for running experiments. Users can avoid typing the commands by copying and pasting them in the code blocks. Inline code elements can be used when introducing the details of code repositories, e.g., configuration parameters, APIs and so on. A README file with more inline code elements is likely to introduce more details about the repository. Hence, we expect that repositories with more code blocks and inline code elements are more likely to be popular. Let us denote the number of back ticks and the number of lines with three back ticks in the README file (in Markdown format) as $bt$ and $bt_3$. The features \emph{num-code-blk} and \emph{num-inline-code} are calculated as $bt_3 / 2$ and $(bt - bt_3 \times 3)/2$, respectively.

Researchers can use images (including animated images), videos and tables to present their paper results. The features \emph{num-img}, \emph{num-ani-img}, \emph{has-video} and \emph{num-table} characterize how researchers present their paper results in the README file. We expect that academic AI repositories showing more results are more likely to be attractive to users and other researchers. For the converted HTML file from the README file, we can extract all the images by looking for the ``$\langle$img$\rangle$'' tag. The \emph{num-img} feature is calculated as the number of images including static and animated images with extensions such as ``jpg'', ``gif''. And the feature \emph{num-ani-img} is calculated as number of animated images with a extension ``gif''. If one of the sentences in $S_{url\_st}$ contains the words ``video'' or ``youtube'', we set \emph{has-video} to 1. We observe that in the HTML file that is converted from the README file, each table starts with a ``$\langle$center$\rangle$'' tag. Hence, the \emph{num-table} feature is calculated as the number of ``$\langle$center$\rangle$'' tags.

Developers recommend that README files should provide links to further documentation or support channels~\citep{prana2018categorizing}. The \emph{num-ghlink} feature refers to the number of links to other GitHub repositories, while the \emph{has-project-page} feature represents whether the README file contains a link to a project page. We observe that the linked GitHub repositories in an academic AI repository can be: 1) the frameworks used by the academic AI repository; 2) the repositories referenced by the academic AI repository; 3) third-party implementation of the paper. These repositories can help users understand more details about the code of the academic AI repository. Furthermore, we observe that a project page provides details about the corresponding paper, e.g., the abstract and idea of the paper. It serves as documentation for introducing the paper rather than specialized documentation of the academic AI repository that corresponds to the paper. Nevertheless, since the project page provides more information about the paper, providing the link to the project page in the README file helps one to understand the repository. Hence, we expect that providing more links to other relevant GitHub repositories and the link to the project page can help the repository attract more users. To calculate \emph{num-ghlink}, we count the number of URLs containing ``github.com'' in the URL set (i.e., $S_{url}$) of the README file. And if one of the sentences in $S_{url\_st}$ of the README file contain ``project'', we set \emph{has-project-page} to 1.

The \emph{has-license} feature represents whether an academic AI repository has a license. A software license is a legal document for guiding the use or redistribution of the software. Users can clearly know how to use the code of a repository that has a license. Hence, we expect that an academic AI repository with a license is more attractive. The license information of a repository is included in its meta-data information. If meta-data information of a repository shows that it has a license, we set \emph{has-license} to 1.

\section{Experiment Results}\label{result}

In this section, we present our analysis respect to the two research questions.

\vspace{0.2cm}\noindent {\bf (RQ1) Is there a relationship between each feature and the popularity of academic AI repositories?}

\vspace{0.1cm}\noindent\textbf{Motivation:} Prior work has investigated the relationship between factors such as documentation and popularity of GitHub repositories~\citep{weber2014makes, zhu2014patterns, aggarwal2014co, borges2016understanding}. To characterize academic AI repositories, we have proposed additional domain-specific features. In this study, we investigate the relationship between each of our features and the popularity of academic AI repositories. AI Researchers can use our results to understand the difference between popular and unpopular academic AI repositories. Based on this understanding, AI researchers can further improve their own academic AI repositories.

\vspace{0.1cm}\noindent\textbf{Method:} For each of the 19 numeric and boolean features (except \emph{language} and \emph{framework}), we apply the Wilcoxon rank-sum test~\citep{wilcoxon1945individual} to analyze statistical significance of difference in means of these feature values between popular and unpopular academic AI repositories. We also calculate Cliff's $\delta$~\citep{cliff2014ordinal} to examine effect size of the difference between the two groups of academic AI repositories. A positive Cliff's $\delta$ indicates that popular academic AI repositories have higher values with respect to the feature than unpopular ones. And a negative Cliff's $\delta$ indicates that popular academic AI repositories have lower values with respect to the feature than unpopular ones. Note that a Cliff's $|\delta|$ less than 0.147, between 0.147 and 0.33, between 0.33 and 0.474, and larger than 0.474 is considered as a negligible (N), small (S), medium (M) and large (L) effect size, respectively~\citep{cliff2014ordinal}.

On the other hand, the features \emph{language} and \emph{framework} are not numeric nor boolean. For popular and unpopular academic AI repositories, we separately calculate the percentage of repositories that use different programming languages and frameworks. For each programming language and machine learning framework, we compare the percentage of popular and unpopular academic AI repositories that use the language or framework. Furthermore, for each language and framework, we calculate the size of four groups of academic AI repositories: (1) popular ones that use the language or framework, (2) popular ones that do not use the language or framework, (3) unpopular ones that use the language or framework and (4) unpopular ones that do not use the language or framework. Then, we apply Fisher's Exact test~\citep{upton1992fisher} on the four numbers to investigate if there is a significant association between using a programming language or framework and being a popular academic AI repository.
Fisher's Exact test is commonly used to determine if there is a significant association between two categorical variables~\citep{fan2018early,wan2018perceptions,xia2019practitioners}.

\vspace{0.1cm}\noindent\textbf{Result:} Table~\ref{tab:relationship} shows the p-values and Cliff's $\delta$ for comparing values of each feature between popular and unpopular academic AI repositories. We observe that the two groups of academic AI repositories have statistically significant difference ($p-value < 0.05$) with a non-negligible effect size for 11 out of the 19 features. For three features, the effect sizes are medium, and for eight features, the effect sizes are small. Hence, popular academic AI repositories have different software engineering practices compared to unpopular ones.

\begin{table}
  \centering
  \scriptsize
  \caption{P-values and Cliff's $\delta$ for the features. Features for which popular and unpopular academic AI repositories show statistically significant difference are in bold.}
      \begin{tabular}{|c|l|r|r|}
    \hline
    \textbf{Dimension} & \textbf{Feature} & \textbf{P-value} & \textbf{Cliff's $\delta$} \\
    \hline
    \multirow{5}{*}{Code} & \textbf{\bf num-code-file} & $<$\textbf{0.001} & \textbf{0.22 (S)}\\
         & \textbf{num-code-line} & $<$\textbf{0.001} & \textbf{0.21 (S)}\\
         & \textbf{num-module} & $<$\textbf{0.001} & \textbf{0.27 (S)}\\
         & num-root-file & $<$0.01 & 0.12 (N)\\
         & num-copy-paste & $<$0.01 & 0.13 (N)\\
    \hline
    \multirow{4}{*}{Reproducibility} & has-data & $<$0.01 & 0.10 (N)\\
                    & has-docker & $<$0.001 & 0.09 (N)\\
                    & \textbf{has-model} & $<$\textbf{0.001} & \textbf{0.28 (S)} \\
                    & \textbf{has-shell} & $<$\textbf{0.001} & \textbf{0.24 (S)} \\
    \hline
    \multirow{10}{*}{Documentation} & \textbf{num-list} & $<$\textbf{0.001} & \textbf{0.38 (M)} \\
                  & \textbf{num-code-blk} & $<$\textbf{0.001} & \textbf{0.35 (M)} \\
                  & \textbf{num-inline-code} & $<$\textbf{0.001} & \textbf{0.30 (S)} \\
                  & \textbf{num-img} & $<$\textbf{0.001} & \textbf{0.36 (M)} \\
                  & num-ani-img & $<$0.001 & 0.09 (N) \\
                  & has-video & $<$0.001 & 0.08 (N) \\
                  & num-table & $<$0.05 & 0.02 (N) \\
                  & \textbf{num-ghlink} & $<$\textbf{0.001} & \textbf{0.31 (S)} \\
                  & has-project-page & $<$0.01 & 0.10 (N) \\
                  & \textbf{has-license} & $<$\textbf{0.001} & \textbf{0.30 (S)}\\
    \hline
    \end{tabular}%
    \vspace{-0.2cm}
  \label{tab:relationship}%
\end{table}%

Now, we analyze each dimension one by one:

For the code dimension, \emph{num-code-file}, \emph{num-code-line} and \emph{num-module} can differentiate popular and unpopular academic AI repositories. Statistical tests show that unpopular academic AI repositories have fewer code files, code lines and modules than popular ones---indicating that unpopular academic AI repositories might not present code detailed enough or be well-modularized. For instance, the \texttt{zole} repository\footnote{https://github.com/Artifineuro/zole} has less than 10 stars (when we collect the data on April 2019). The repository only has 1 code file containing 32 lines of code. It does not provide the code for training models. And another example is the \texttt{EndToEndIncrementalLearning} repository,\footnote{https://github.com/fmcp/EndToEndIncrementalLearning} which has less than 30 stars (as of April 2019). The repository is not well-modularized: it directly presents many code files in the root directory without arranging them into sub-directories (modules).

For the reproducibility dimension, \emph{has-model} and \emph{has-shell} can differentiate popular and unpopular academic AI repositories. Statistical tests showing that popular academic AI repositories tend to provide a pre-trained model and shell-scripts in comparison with unpopular academic AI repositories---indicating that popular academic AI repositories tend to take more measures to ease the reproducibility of the corresponding papers. For instance, the \texttt{StarGAN} repository\footnote{https://github.com/yunjey/stargan} has more than 3,000 stars. The repository provides pre-trained models and a shell script for downloading the models. The models and script make it easier for others to use the repository.

For the documentation dimension, \emph{num-list}, \emph{num-code-blk}, \emph{num-inline-code}, \emph{num-img}, \emph{num-ghlink} and \emph{has-license} can differentiate popular and unpopular academic AI repositories. Statistical tests suggest that popular academic AI repositories tend to use more lists, code blocks, inline code elements and images to organize the contents of README compared to unpopular ones. Popular academic AI repositories tend to have more links to other GitHub repositories than unpopular ones. Statistical tests also suggest that popular academic AI repositories tend to have a license compared to unpopular ones. Hence, the documentation of popular academic AI repositories tends to have better quality. For instance, the \texttt{vid2vid} repository\footnote{https://github.com/NVIDIA/vid2vid} has more than 6,000 stars. The README file of the repository contains 26 lists, 16 code blocks, 80 inline code elements, 7 images and 4 links to other GitHub repositories. Moreover, the repository has a license.

\begin{table*}
  \centering
  \scriptsize
  \caption{Percentage of repositories that use different programming languages in popular academic AI repositories as compared to unpopular ones. We also show p-values that are calculated using Fisher's Exact test. Languages that are significantly associated with popularity of academic AI repositories are in bold.}
    \begin{tabular}{|l|r|r|r|r|r|}
    \hline
    \textbf{Language} & \textbf{C/C++} & \textbf{Lua} & \textbf{MATLAB} & \textbf{Python} & \textbf{Other} \\
    \hline
    Popular & 12.0\% & 10.3\% & \textbf{12.9\%} & 63.5\% & 1.3\% \\
    Unpopular & 10.2\% & 7.4\% & \textbf{19.7\%} & 60.1\% & 2.6\% \\
    \hline
    { P-value} & $>$0.05 & $>$0.05 & \textbf{$<$0.05} & $>$0.05 & $>$0.05 \\
    \hline
    \end{tabular}%
  \label{tab:language}%
\end{table*}%

\begin{table*}
  \centering
  \scriptsize
  \caption{Percentage of repositories that use different frameworks in popular academic AI repositories as compared to unpopular ones. We also show the p-values that are calculated using Fisher's Exact test. Frameworks that are significantly associated with popularity of academic AI repositories are in bold.}
    \resizebox{1.1\textwidth}{!}{
    \begin{tabular}{|l|r|r|r|r|r|r|r|r|}
    \hline
    \textbf{Framework} & \textbf{Caffe} & \textbf{MATLAB} & \textbf{PyTorch} & \textbf{Tensorflow} & \textbf{Torch} & \textbf{MXNet} & \textbf{Other} & \textbf{Unknown} \\
    \hline
    Popular & 15.5\% & \textbf{12.9\%} & \textbf{26.2\%} & 24.5\% & \textbf{10.7\%} & \textbf{3.0\%} & 2.0\% & \textbf{5.2\%} \\
    Unpopular & 11.4\% & \textbf{19.7\%} & \textbf{13.8\%} & 23.4\% & \textbf{6.4\%} & \textbf{0.6\%} & 1.2\% & \textbf{23.5\%} \\
    \hline
    {P-value} & $>$0.05 & \textbf{$<$0.05} & \textbf{$<$0.001} & $>$0.05 & \textbf{$<$0.05} & \textbf{$<$0.01} & $>$0.05 & \textbf{$<$0.001}\\
    \hline
    \end{tabular}%
    }
  \label{tab:framework}%
\end{table*}%

Table~\ref{tab:language} and Table~\ref{tab:framework} present the percentage of repositories that use different programming languages and frameworks in popular academic AI repositories compared to unpopular ones, respectively. Notice that we only present the details of programming languages and frameworks that are used by at least 10 academic AI repositories. For other programming languages and frameworks, we summarize the percentages of these programming languages or frameworks in a category named \emph{Other}. Furthermore, in Table~\ref{tab:language}, we present the p-values for comparing popular and unpopular academic AI repositories that use and do not use a programming language. And in Table~\ref{tab:framework}, we also show the p-values for comparing popular and unpopular academic AI repositories that use and do not use a machine learning framework.

From Tables~\ref{tab:language} and~\ref{tab:framework}, we find that there is a significant association between using a programming language or framework and being a popular academic AI repository. Academic AI repositories that use MATLAB are less likely to be popular. MATLAB is a specialized numeric computing environment and does not provide the needed functionality to implement general applications. Hence, it may be difficult to incorporate algorithms implemented using MATLAB into applications that are implemented in other programming languages (e.g., Python). Furthermore, MATLAB is commercial and expensive, which limits its popularity. On the other hand, academic AI repositories that use PyTorch, Torch, and MXNet are more likely to be popular. PyTorch is a machine learning framework based on Python, which is easy to use and consistent with other popular scientific computing libraries~\citep{paszke2019pytorch}. Torch is a scientific computing framework based on Lua, which aims to provide maximum flexibility and speed in building scientific algorithms~\citep{collobert2002torch}. MXNet is a flexible and efficient library for deep learning, and it supports eight programming languages. PyTorch, Torch, and MXNet are all open source and freely available on the web.

Furthermore, we find that a much larger percentage of unpopular academic AI repositories are in the Unknown category---indicating that unpopular academic AI repositories are less likely to explicitly state the used machine learning framework in the README file. Furthermore, Fisher's Exact test shows that there is a significant association between explicitly stating the used framework in the README file and being a popular academic AI repository. This finding indicates that unpopular academic AI repositories are less likely to provide complete information in their documentation compared to popular ones.

\vspace{0.2cm}\noindent \fbox{\parbox[c]{1\textwidth} {\emph{Popular academic AI repositories are statistically significantly different from unpopular academic repositories with a non-negligible effect size in 11 of the 19 numerical features. Generally, popular academic AI repositories have more code files and modules than unpopular ones. They take more measures to ease the reproducibility of their experiments. And their documentation has better quality. Furthermore, academic AI repositories that use MATLAB and those that do not explicitly state the used framework in the README file are less likely to be popular. Academic AI repositories that use PyTorch, Torch, and MXNet are more likely to be popular.}}}

\vspace{0.2cm}\noindent {\bf (RQ2) What are the most important features that differentiate popular and unpopular academic AI repositories?}

\vspace{0.1cm}\noindent\textbf{Motivation:} Among our repository features, some features may be more important than the others in differentiating popular and unpopular academic AI repositories. In this research question, we investigate the most important features in differentiating the two groups of repositories. Then researchers can pay more attention to these important features to improve their academic AI repositories.

\vspace{0.1cm}\noindent\textbf{Method:} To compare the importance of the studied features, we build a random forest classifier~\citep{breiman2001random} to discriminate whether a repository is popular, given values of our repository features. The features that contribute more to the classifier is considered to be more important.

Prior to building a model, we first remove correlated features since correlated features can impact model interpretation~\citep{tantithamthavorn2018experience}. Following Harrel's guidelines for building models~\citep{harrell2015regression}, we remove correlated features using the following three steps:

\vspace{0.05cm}\noindent\underline{\em Step 1 Collinearity Analysis.} In this step, we mitigate the correlation between each pair of features. We calculate Spearman's correlation coefficient ($\rho$) between each pair of the repository features. We set 0.7 as the correlation threshold. If two features has a $|\rho|$ larger than 0.7, we randomly drop one of the features. As a result, we find that \emph{num-code-file} and \emph{num-code-line} are highly correlated and \emph{num-code-line} is dropped.

\vspace{0.05cm}\noindent\underline{\em Step 2 Independence Analysis.} Note that Spearman's correlation coefficient cannot be applied on non-numerical features, i.e., \emph{framework} and \emph{language}. In this step, we examine whether the two non-numerical features are independent of the other numerical features. For the two features, we use Chi-squared test of independence to analyze whether each of the two features are statistically independent from the other features. We use the chisq.test function in the \textbf{vcd} R package.\footnote{https://cran.r-project.org/web/packages/vcd/index.html} As a result, we find that neither of the two features is independent from the other features (e.g., \emph{num-code-file}), and both features are dropped.

\vspace{0.05cm}\noindent\underline{\em Step 3 Redundancy Analysis.} In this step, we remove redundant features that can be predicted by combining other features, since redundant features cannot further improve model performance in the appearance of other features. We apply the redun function in the \textbf{rms} R package on the remaining features.\footnote{https://cran.r-project.org/web/packages/rms/index.html} As a result, no redundant feature is detected.

After the above three steps, we remove three features, i.e., \emph{num-code-line}, \emph{framework}, and \emph{language}. For the remaining 18 features, we calculate the most important features for differentiating popular and unpopular academic AI repositories using 1,000 times of out-of-sample bootstrap following \citet{tantithamthavorn2017empirical}. We leverage a random forest classifier~\citep{breiman2001random} to analyze the importance of the 18 features. Random forest is an ensemble approach and specially designed for decision tree classifiers. It is composed of multiple decision tree classifiers. To label a sample, a random forest classifier applies the mode of the labels that are output by individual decision trees. In comparison to other classifiers, random forest is generally highly accurate~\citep{yan2018automating,tantithamthavorn2018impact,fan2018early,fan2019impact, ghotra2015revisiting}. Random forest is also robust to outliers since it summarizes the classification results of multiple trees that are learned differently~\citep{cutler2012random}. We use the \textbf{randomForest} R package to build the random forest classifier.\footnote{https://cran.r-project.org/web/packages/randomForest/index.html}

In each bootstrap iteration, a training set is generated by randomly sampling 1,040 repositories with replacement from our repository data set, and the repositories that are not sampled are used as the testing set. We first build a random forest classifier on the training set. Then, we apply the classifier on the testing set and calculate AUC~\citep{huang2005using} to investigate whether the classifier has good discrimination power. And we calculate~\cite{tantithamthavorn2018impact}'s importance score for the 18 features. We take two steps to calculate the importance score for each feature. First, we permutate the feature values in the testing set, producing a new testing set with the feature permutated and the other features as is. Second, we compute the difference of the misclassification rates when the classifier is applied on the original testing set and when the classifier is applied on the testing set with the feature permutated. A larger difference indicates that the feature is more important. Therefore, the value of the difference is considered as the importance score of the feature.

After the 1,000 times of out-of-sample bootstrap, we have 1,000 AUC scores of the random forest classifier. We calculate the average AUC score. Furthermore, using the 1,000 importance scores for each feature from the 1,000 iterations of our out-of-sample bootstrap, we apply the Scott-Knott Effect Size Difference (SK-ESD) test~\citep{tantithamthavorn2017empirical, fan2018chaff}. The SK-ESD test is an enhancement of the Scott-Knott test~\citep{scott1974cluster}. By mitigating the skewness of input data, the SK-ESD test relaxes the assumption of normally distributed data (which is required by the Scott-Knott test). Furthermore, the SK-ESD test also considers the effect size of input data and merges any two distinct groups when the groups have a negligible effect size.

\vspace{0.1cm}\noindent\textbf{Result:} We find that the average AUC score of the classifier across the 1,000 times of out-of-sample bootstrap is 0.79. An AUC larger than 0.7 is generally considered as acceptable performance~\citep{hosmer2013applied,fan2018early}. This result confirms that our repository features can effectively differentiate popular and unpopular academic AI repositories.

\begin{table}
  \centering
  \caption{The SK-ESD test results when comparing the rank of feature importance. The features are divided into distinct groups that have statistically significant difference.}
    \begin{tabular}{|c|l|l|}
    \hline
    \textbf{Group} & \textbf{Features} & \textbf{Dimension} \\
    \hline
    \multirow{2}{*}{G1} & num-ghlink & Documentation \\
    & num-img & Documentation \\
    \hline
    G2 & has-license & Documentation \\
    \hline
    \multirow{3}{*}{G3} & has-shell & Reproducibility \\
    & num-code-blk & Documentation \\
    & num-list & Documentation \\
    \hline
    \multirow{2}{*}{G4} & num-inline-code & Documentation \\
    & num-root-file & Code \\
    \hline
    \multirow{5}{*}{G5} & num-code-file & Code \\
    & has-project-page & Documentation \\
    & has-video & Documentation \\
    & has-docker & Reproducibility \\
    & has-trained-model & Reproducibility \\
    \hline
    \multirow{5}{*}{G6} & num-module & Code \\
    & num-table & Documentation \\
    & has-data & Reproducibility \\
    & num-copy-paste & Code \\
    & num-ani-image & Documentation\\
    \hline
    \end{tabular}%
    \vspace{-0.3cm}
  \label{tab:rank}%

\end{table}%

Table~\ref{tab:rank} presents the 18 features ranked by the SK-ESD test. Different groups of features have statistically significant difference in terms of feature importance. As shown in Table~\ref{tab:rank}, \emph{num-ghlink}, \emph{num-img} and \emph{has-license} are the top-3 important features for differentiating popular and unpopular academic AI repositories. From Table~\ref{tab:relationship}, we notice that popular and unpopular academic AI repositories are statistically significantly different with respect to the top-3 important features. And the effect sizes of the three features are all positive and non-negligible. Hence, \emph{num-ghlink}, \emph{num-img} and \emph{has-license} are the most important features that are positively associated with the popularity of academic AI repositories. Since all of the three features are from the documentation dimension, AI researchers should pay more attention to the documentation factors of their academic AI repositories.

\vspace{0.2cm}\noindent \fbox{\parbox[c]{1\textwidth} {\emph{The number of links to other GitHub repositories in the README file, the number of images in the README file and inclusion of a license are the most important factors to distinguish popular and unpopular academic AI repositories.}}}

\section{Discussion}\label{discuss2}


\subsection{External contributions to academic AI repositories}\label{dis:other-metric}

In GitHub, a user can fork a repository, i.e., make a personal copy of the repository with the permission of the repository owner.~\cite{jiang2017and} observed that forking is mainly used for making contributions to original repositories. The user can make changes to the forked repository and submit pull requests to integrate the changes back to the original repository~\citep{gousios2014exploratory}. In this study, we consider a developer as an external contributor of an academic AI repository if he/she is not the repository owner and submits pull requests to the repository. We use the number of forks, pull requests, and external contributors to measure external contributions to the studied academic AI repositories. Notice that when counting number of pull requests for an academic AI repository, we do not consider the pull requests submitted by the repository owner.

In this study, we use the number of stars as a proxy for measuring the popularity of an academic AI repository. This popularity measure is commonly used by prior studies~\citep{borges2016predicting,borges2016understanding,hu2016influence}. In our preliminary study, we show that citations of AI papers and the number of stars of the papers' accompanying repositories are strongly correlated. In addition to the number of stars, contribution-based measures can also reflect the popularity of a GitHub repository. In this section, we investigate the correlation between citations of papers and the contribution-based measures of their accompanying repositories. Furthermore, we also investigate the correlation between the number of stars and the three contribution-based measures for the studied academic AI repositories.

For each of the 1,149 studied academic AI repositories, we retrieve the number of forks and all of the pull requests via the GitHub REST API.\footnote{https://developer.github.com/v3/} Then, we calculate external contributors of the academic AI repository by extracting the owner of each pull request. Notice that we do not consider the owner of a repository as an external contributor of the repository. Also, we do not consider the pull requests submitted by the repository owner. For each studied academic AI repository, we calculate the number of forks, pull requests, and external contributors.

Similarly to Section~\ref{sec:preliminary}, for papers published in each year, we separately calculate the correlation between citations of papers and the three contribution-based measures of their accompanying repositories. We calculate the correlation using the Spearman rank correlation test. Also, we drop the papers published in 2018 since they may not get enough citations. Furthermore, we use all of the 1,149 academic AI repositories to analyze the correlation between the number of stars and the three contribution-based measures. And we calculate the correlation using the Spearman rank correlation test.

\begin{table*}
  \centering
  \scriptsize
  \caption{Correlation between citations of papers and number of forks, pull requests and external contributors of the papers' accompanying repositories, respectively.}
    \begin{tabular}{|l|r|r|r|}
    \hline
    \textbf{Measures} & \multicolumn{1}{l|}{\textbf{Years}} & \multicolumn{1}{l|}{\textbf{Spearman's $\rho$}} & \textbf{P-value} \\
    \hline
    \hline
    \multirow{3}[0]{*}{\textbf{\#fork}} & 2017  & 0.69  & $<$0.001 \\
          & 2016  & 0.69  & $<$0.001 \\
          & 2015  & 0.75  & $<$0.001 \\
    \hline
    \hline
    \multicolumn{1}{|l|}{\multirow{3}[0]{*}{\textbf{\#pull request}}} & 2017  & 0.33  & $<$0.001 \\
          & 2016  & 0.41  & $<$0.001 \\
          & 2015  & 0.37  & $<$0.01 \\
    \hline
    \hline
    \multirow{3}[0]{*}{\textbf{\#contributor}} & 2017  & 0.34  & $<$0.001 \\
          & 2016  & 0.42  & $<$0.001 \\
          & 2015  & 0.38  & $<$0.01 \\
    \hline
    \end{tabular}%
  \label{tab:contrib-cor}%
\end{table*}%

\begin{table*}
  \centering
  \scriptsize
  \caption{Correlation between number of stars and number of forks, pull requests and external contributors of academic AI repositories, respectively.}
    \begin{tabular}{|l|r|l|}
    \hline
    \textbf{Mesures} & \multicolumn{1}{l}{\textbf{Spearman's $\rho$}} & \textbf{P-value} \\
    \hline
    \textbf{\#fork} & 0.92  & $<$0.001 \\
    \textbf{\#pull request} & 0.50  & $<$0.001 \\
    \textbf{\#contributor} & 0.50  & $<$0.001 \\
    \hline
    \end{tabular}%
  \label{tab:cor-star}%
\end{table*}%

Table~\ref{tab:contrib-cor} presents the correlation between citations of papers and the three contribution-based measures of their accompanying academic AI repositories, respectively. As shown in the table, the correlation coefficient between the citations of papers and the number of forks of their accompanying repositories is still close to or above 0.7. The p-value is less than 0.001. Thus, there is still a relatively high correlation between the paper citation and the number of forks of the accompanying repository. On the other hand, we find that for papers published in each year, the paper citation is weakly correlated with the number of pull requests and the number of external contributors of the accompanying academic AI repositories with a Spearman's $\rho$ ranging from 0.34 to 0.42 and a p-value of less than 0.01.

Table~\ref{tab:cor-star} presents the correlation between the number of stars and the three contribution-based measures of the studied academic AI repositories, respectively. From the table, we find that for academic AI repositories, the number of stars and the number of forks are very strongly correlated (with a Spearman's $\rho > 0.9$ and $p-value < 0.001$). This finding is consistent with prior work~\citep{borges2016understanding}. On the other hand, we find that for academic AI repositories, the number of stars is moderately correlated with the number of pull requests and the number of external contributors with a Spearman's $\rho$ of 0.50 and a p-value of less than 0.001.

We observe that in comparison to stars and forks, academic AI repositories have a limited number of external contributors and pull requests. On average, the 1,149 academic AI repositories have 163.9 stars, 37.0 forks, 1.6 pull requests, and 0.7 contributors. We find that 864 academic AI repositories do not have external contributors submitting pull requests. These academic AI repositories have no pull requests and external contributors. But they have different numbers of stars, and their corresponding papers have different citations, which explains why the number of pull requests and the number of external contributors of the academic AI repositories are not highly correlated with the number of stars citations of corresponding papers.

In GitHub, starring and forking a repository can be completed by simply pushing buttons, while submitting pull requests requires making changes to code of the repository. Contributors of an academic AI repository may face challenges because making changes to such a repository requires AI knowledge and comprehension of the corresponding paper. We randomly select 50 external contributors of the academic AI repositories in our dataset. From profiles of the contributors on GitHub, we find that 22 of the contributors are from academia, and 14 of the contributors are from industry. For the other contributors, their profiles do not provide enough information for us to identify their work. Furthermore, we find that 2 of the industrial contributors work with AI. Thus, academic AI repositories may have requirements for contributors in AI knowledge and the ability to understand the corresponding papers.

\vspace{0.2cm}\noindent \fbox{\parbox[c]{1\textwidth} {\emph{Paper citation is highly correlated with the number of forks of the accompanying academic AI repositories and weakly correlated with the number of pull requests and external contributors of the accompanying academic AI repositories. For academic AI repositories, the number of stars is very strongly correlated with the number of forks and moderately correlated with the number of pull requests and external contributors. In comparison to stars and forks, academic AI repositories have a limited number of pull requests and external contributors, since academic AI repositories require contributors to possess sufficient AI knowledge and the ability to understand the corresponding papers.}}}

\subsection{Differences between personal and organizational academic AI repositories}\label{dis:owner}

A GitHub repository can be owned by a personal user or an organization~\citep{borges2016understanding}. The meta-data information of an academic AI repository that we crawled contains an ``organization'' field, which can indicate if the repository owner is an organization. In this section, we investigate the extent to which the repository owner is associated with the popularity of academic AI repositories. Furthermore, we compare personal and organizational repositories. We investigate whether they are different concerning the relationship between software engineering practices and the popularity of repositories. By doing so, we may derive more suggestions that are specific for personal and organizational repositories.

We first investigate whether the repository owner is associated with the popularity of academic AI repositories. Among the labeled 1,040 academic AI repositories, we find that 891 academic AI repositories are owned by personal users, in which 180 repositories are popular. And we find that 149 academic AI repositories are owned by organizations, in which 53 repositories are popular. We calculate the number of popular and unpopular academic AI repositories that are owned by personal users and organizations, respectively. Similarly to RQ1, we apply Fisher's Exact test on the calculated numbers. Fisher's Exact test shows that there is a significant association between the owner of an academic AI repository and the popularity of the repository with a p-value of less than 0.001. Organizational academic AI repositories are more likely to be popular than personal ones. This finding is consistent with prior work~\citep{borges2016understanding}.

Then, we investigate whether personal and organizational repositories are different concerning the relationship between software engineering practices and the popularity of repositories. We apply the described method in RQ1 to separately analyze the relationship between each feature and the repository popularity for personal and organizational repositories. For the two groups of repositories, we apply the Wilcoxon rank-sum test and calculate Cliff's $\delta$ for each of the 19 numeric and boolean features. Also, for the two groups of repositories, we separately calculate the percentage of repositories using each programming language and framework in popular and unpopular academic AI repositories. Also, we apply Fisher's Exact test to investigate if using a language or framework is associated with the popularity of academic AI repositories.

\begin{table*}
  \centering
  \scriptsize
  \caption{P-values and Cliff's $\delta$ for the features of popular academic AI repositories as compared to unpopular ones. We separately show results for personal and organizational repositories. Features for which popular and unpopular academic AI repositories show statistically significant difference are in bold.}
    \begin{tabular}{|c|l|r|r|r|r|}
    \hline
    \multirow{2}[0]{*}{\textbf{Dimension}} & \multicolumn{1}{c|}{\multirow{2}[0]{*}{\textbf{Feature}}} & \multicolumn{2}{c|}{\textbf{Personal}} & \multicolumn{2}{c|}{\textbf{Organizational}} \\
    \cline{3-6}
          &       & \textbf{P-value} & \textbf{Cliff's $\delta$} & \textbf{P-value} & \textbf{Cliff's $\delta$} \\
    \hline
    \multirow{5}[0]{*}{Code} & num-code-file & \textbf{$<$0.001} & \textbf{0.19 (S)} & \textbf{$<$0.001} & \textbf{0.34 (M) } \\
          & num-code-line & \textbf{$<$0.001} & \textbf{0.18 (S)} & \textbf{$<$0.01} & \textbf{0.30 (S)} \\
          & num-module & \textbf{$<$0.001} & \textbf{0.27 (S)} & \textbf{$<$0.01} & \textbf{0.28 (S)} \\
          & num-root-file & $<$0.01 & 0.12 (N) & $>$0.05 & 0.11 (N) \\
          & num-copy-paste & $<$0.05 & 0.11 (N) & \textbf{$<$0.01} & \textbf{0.26 (S)} \\
    \hline
    \multirow{4}[0]{*}{Reproducibility} & has-data & $<$0.05 & 0.08 (N) & \textbf{$<$0.05} & \textbf{0.20 (S)} \\
          & has-docker & $<$0.001 & 0.08 (N) & $<$0.05 & 0.12 (N) \\
          & has-model & \textbf{$<$0.001} & \textbf{0.27 (S)} & \textbf{$<$0.001} & \textbf{0.35 (M)} \\
          & has-shell & \textbf{$<$0.001} & \textbf{0.21 (S)} & \textbf{$<$0.001} & \textbf{0.37 (M)} \\
    \hline
    \multirow{10}[0]{*}{Documentation} & num-list & \textbf{$<$0.001} & \textbf{0.40 (M)} & \textbf{$<$0.01} & \textbf{0.30 (S)} \\
          & num-code-blk & \textbf{$<$0.001} & \textbf{0.33 (M)} & \textbf{$<$0.001} & \textbf{0.40 (M)} \\
          & num-inline-code & \textbf{$<$0.001} & \textbf{0.27 (S)} & \textbf{$<$0.001} & \textbf{0.33 (S)} \\
          & num-img & \textbf{$<$0.001} & \textbf{0.35 (M)} & \textbf{$<$0.001} & \textbf{0.43 (M)} \\
          & num-ani-img & $<$0.01 & 0.06 (N) & \textbf{$<$0.01} & \textbf{0.16 (S)} \\
          & has-video & $<$0.001 & 0.08 (N) & $>$0.05 & 0.10 (N) \\
          & num-table & $<$0.05 & 0.02 (N) & $>$0.05 & 0.01 (N) \\
          & num-ghlink & \textbf{$<$0.001} & \textbf{0.30 (S)} & \textbf{$<$0.001} & \textbf{0.34 (M)} \\
          & has-project-page & $<$0.001 & 0.13 (N) & $>$0.05 & -0.01 (N) \\
          & has-license & \textbf{$<$0.001} & \textbf{0.28 (S)} & \textbf{$<$0.001} & \textbf{ 0.23 (S)} \\
    \hline
    \end{tabular}%
  \label{tab:owner-feature}%
\end{table*}%

For personal and organizational repositories, Table~\ref{tab:owner-feature} separately shows the p-values and Cliff's $\delta$ for comparing values of each feature between popular and unpopular repositories. For personal repositories, we find that popular and unpopular repositories have a statistically significant difference ($p-value < 0.05$) with a non-negligible effect size for 11 out of the 19 features. For two features, the effect sizes are medium, and for nine features, the effect sizes are small. For organizational repositories, we find that popular and unpopular repositories have a statistically significant difference ($p-value < 0.05$) with a non-negligible effect size for 14 of the 19 features. For six features, the effect sizes are medium, and for eight features, the effect sizes are small.

We compare the results shown in Table~\ref{tab:owner-feature} and Table~\ref{tab:relationship}. In comparison to all the labeled academic AI repositories, personal and organizational repositories have the same 11 features that show a significant difference between popular and unpopular repositories. Personal repositories do not have additional features that show a significant difference between popular and unpopular repositories. Organizational repositories have another three features that show a significant difference between popular and unpopular repositories, namely \emph{num-copy-paste}, \emph{has-data}, and \emph{num-ani-img}.

For the \emph{num-copy-paste} feature, statistical tests show that among organizational repositories, popular repositories more frequently copy and paste code than unpopular ones. According to Table~\ref{tab:owner-feature}, we notice that among organizational repositories, popular repositories have significantly more code lines than unpopular ones. We hypothesize that there is a positive correlation between the scale of code and frequency of copying and pasting code, which may result in that popular repositories more frequently copy and paste code than unpopular ones. For organizational repositories, we calculate the correlation between values of \emph{num-copy-paste} and \emph{num-code-line} using the Spearman rank correlation test. The calculated Spearman's $\rho$ is 0.59, showing that the two features are moderately and positively correlated~\citep{schober2018correlation}. Furthermore, we leverage the \emph{num-code-line} feature to normalize the \emph{num-copy-paste} feature. We apply the Wilcoxon rank-sum test to compare the normalized \emph{num-copy-paste} feature between popular and unpopular academic AI repositories that are created by organizations. The statistical test shows that the two groups of repositories do not show a significant difference ($p-value > 0.05$) in terms of the normalized \emph{num-copy-paste} feature. Thus, for organizational repositories, we do not suggest to more frequently copy and paste code.

For the \emph{has-data} feature, statistical tests show that for organizational repositories, popular repositories tend to provide datasets in comparison with unpopular ones. For the \emph{num-ani-img} feature, statistical tests show that for organizational repositories, popular repositories present more animated images in the README file as compared to unpopular ones. Thus, we suggest that organizational repositories should provide datasets and present more animated images in the README file.

\begin{table*}
  \centering
  \scriptsize
  \caption{Percentage of repositories that use different programming languages in popular academic AI repositories as compared to unpopular ones. We separately show the results for personal and organizational repositories. We also show p-values calculated using Fisher's Exact test.}
    \begin{tabular}{|c|l|r|r|r|r|r|}
    \hline
    \textbf{Owner} & \textbf{Popularity} & \multicolumn{1}{l|}{\textbf{C/C++}} & \multicolumn{1}{l|}{\textbf{Lua}} & \multicolumn{1}{l|}{\textbf{MATLAB}} &
    \multicolumn{1}{l|}{\textbf{Python}} & \multicolumn{1}{l|}{\textbf{Other}} \\
    \hline
    \hline
    \multirow{3}[0]{*}{Personal} & Popular & 13.3\% & 10.0\% & 15.0\% & 60.6\% & 1.1\% \\
          & Unpopular & 9.6\% & 7.7\% & 21.4\% & 58.5\% & 2.8\% \\
    \cline{2-7}
          & {P-value} & $>$0.05 & $>$0.05 & $>$0.05 & $>$0.05 & $>$0.05 \\
    \hline
    \hline
    \multirow{3}[0]{*}{Organizational} & Popular & 7.5\% & 11.3\% & 5.7\% & 73.6\% & 1.9\% \\
          & Unpopular & 14.6\% & 5.2\% & 7.3\% & 71.9\% & 1.0\% \\
    \cline{2-7}
          & P-value & $>$0.05 & $>$0.05 & $>$0.05 & $>$0.05 & $>$0.05 \\
    \hline
    \end{tabular}%
  \label{tab:owner-lang}%
\end{table*}%

For personal and organizational repositories, Table~\ref{tab:owner-lang} separately presents the percentage of repositories that use different programming languages in popular repositories as compared to unpopular ones. We also show p-values calculated using Fisher's Exact test in the table. In our answer to RQ1, Table~\ref{tab:language} shows that when considering all of the labeled academic AI repositories, usage of MATLAB is significantly associated with the popularity of repositories. However, in Table~\ref{tab:owner-lang}, we find that for personal and  organizational repositories, there is no significant association between using a language and being a popular academic AI repository. We notice that for both groups of repositories, the percentage of unpopular repositories using MATLAB is larger than that of popular ones using MATLAB. Thus, when considering all the labeled repositories, the association between using MATLAB and the popularity of repositories becomes more significant than separately considering personal or organizational repositories.

\begin{table*}
  \centering
  \scriptsize
  \caption{Percentage of repositories that use different frameworks in popular academic AI repositories as compared to unpopular ones. We separately show the results for personal and organizational repositories. We also show p-values calculated using Fisher's Exact test. Frameworks that are significantly associated with popularity of repositories are in bold.}
  \resizebox{1.1\textwidth}{!}{
  \begin{tabular}{|c|l|r|r|r|r|r|r|r|r|}
  \hline
    \multicolumn{1}{|l|}{\textbf{Owner}} & \textbf{Popularity} & \textbf{Caffe} & \textbf{MATLAB} & \textbf{PyTorch} & \textbf{Tensorflow} & \textbf{Torch} & \textbf{MXNet} & \textbf{Other} & \textbf{Unknown} \\
    \hline
    \hline
    \multirow{3}[0]{*}{Personal} & Popular & 16.7\% & 15.0\% & \textbf{26.1\%} & 26.7\% & 10.0\% & 1.1\% & 0.6\% & \textbf{3.8\%} \\
          & Unpopular & 11.8\% & 21.4\% & \textbf{14.2\%} & 21.4\% & 6.6\% & 0.4\% & 0.8\% & \textbf{23.4\%} \\
    \cline{2-10}
          & {P-value} & $>$0.05 & $>$0.05 & \textbf{$<$0.001} & $>$0.05 & $>$0.05 & $>$0.05 & $>$0.05 & \textbf{$<$0.001} \\
    \hline
    \hline
    \multirow{3}[0]{*}{Organizational} & Popular & 11.3\% & 5.7\% & \textbf{26.4\%} & \textbf{17.0\%} & 13.2\% & 9.4\% & 7.6\% & \textbf{9.4\%} \\
          & Unpopular & 8.3\% & 7.3\% & \textbf{10.4\%} & \textbf{38.5\%} & 5.2\% & 2.1\% & 3.10\% & \textbf{25.1\%} \\
    \cline{2-10}
          & {P-value} & $>$0.05 & $>$0.05 & \textbf{$<$0.05} & \textbf{$<$0.01} & $>$0.05 & $>$0.05 & $>$0.05 & \textbf{$<$0.05} \\
    \hline
    \end{tabular}%
    }
  \label{tab:owner-framework}%
\end{table*}%

For personal and organizational repositories, Table~\ref{tab:owner-framework} separately shows the percentage of repositories that use different machine learning frameworks in popular repositories as compared to unpopular ones. We also show p-values calculated using Fisher's Exact test in the table. For both personal and organizational repositories, we find that there is a significant association between using a framework and being a popular academic AI repository. In comparison to our findings from Table~\ref{tab:framework} in our answer to RQ1, we have similar findings for both groups of repositories: (1) academic AI repositories using PyTorch are more likely to be popular. (2) academic AI repositories that do not explicitly state the used framework in the README file are less likely to be popular. Besides, Table~\ref{tab:owner-framework} shows that for personal or organizational repositories, there is no significant association between using Torch or MXNet and being a popular academic AI repository. From Table~\ref{tab:framework}, we find that when considering all the labeled academic AI repositories, using Torch and MXNet is significantly associated with the popularity of the repositories. For both groups of repositories, we notice that the percentage of popular repositories using Torch and MXNet is larger than that of unpopular repositories using the two frameworks. Thus, when considering all the labeled repositories, the association between using Torch and MXNet and the popularity of repositories becomes more significant than separately considering personal or organizational repositories.

Furthermore, for organizational repositories, Fisher's Exact test shows that there is a significant association between using Tensorflow and being a popular academic AI repository. We notice that organizational repositories that use Tensorflow are less likely to be popular.

\vspace{0.2cm}\noindent \fbox{\parbox[c]{1\textwidth} {\emph{There is a significant association between the owner of an academic AI repository and the popularity of the repository. Personal and organizational repositories are different concerning the relationship between software engineering practices and the popularity of repositories. Among personal repositories, popular and unpopular ones show a significant difference for 11 features. Among organizational repositories, popular and unpopular ones show a significant difference for 14 features. Different from personal repositories, organizational repositories that provide datasets and present more animated images in the README file are more likely to be popular. Furthermore, organizational repositories that use Tensorflow are less likely to be popular.}}}

\subsection{The relationship between paper topics and the popularity of academic AI repositories}\label{dis:topic}

In practice, different people may focus on AI papers on different topics. In this section, we investigate the degree to which topics of AI papers are associated with the popularity of their accompanying academic AI repositories.

Since an abstract serves as the summary of a paper, we consider that the abstract of a paper is enough for us to analyze the topic of the paper. For each labeled academic AI repository, we crawl the corresponding paper and extract the abstract of the paper. As a result, we retrieve 1,040 documents, and each document is the abstract of an AI paper. Leveraging the 1,040 documents, we build a topic model~\citep{blei2003latent} to discover the topics that occur in the 1,040 documents. Given a set of documents, a topic model calculates topics of the documents as clusters of words or n-grams. Each topic is represented as a probability distribution of words or n-grams. And each document can be represented as a probability distribution of the output topics by the topic model.

We first use the Latent Dirichlet Allocation (LDA) technique~\citep{blei2003latent} to build the topic model. We leverage the LDA implementation in the \textbf{Gensim} Python package.\footnote{https://radimrehurek.com/gensim/} Following Prabhakaran's guidelines,\footnote{https://www.machinelearningplus.com/nlp/topic-modeling-gensim-python/} we preprocess the 1,040 documents before we learn the topic model. We first preprocess the documents, including removing stop words and word lemmatization. In addition to uni-grams (i.e., words), we also consider bi-grams since phrases like ``object detection'' may reflect the topic of a paper. As a result, we represent each document as a group of uni-grams and bi-grams. Then, we apply the LDA technique for the documents. We vary the number of topics from 5 to 15 and build multiple topic models. For each topic model, we calculate the topic coherence measure~\citep{newman2010automatic}. Topic coherence measures the quality of a topic model. The key idea of topic coherence is to estimate the degree of semantic similarity between high scoring words or n-grams in each topic~\citep{newman2010automatic}. A higher topic coherence indicates a better topic model. As a result, we find that the topic model with ten topics achieves the highest topic coherence. Hence, we apply this topic model for further analysis. Table~\ref{tab:topic} shows the ten topics of the topic model. Notice that we summarize the eighth topic as ``Reinforcement Learning'' since the keywords ``agent'' and ``policy'' are frequently used in the reinforcement learning area~\citep{sutton2018reinforcement}.

\begin{table*}
	\centering
    \scriptsize
	\caption{Topics of corresponding AI papers of the studied academic AI repositories that are calculated using the LDA technique. We also show p-values and Cliff's $\delta$ for comparing popular and unpopular academic AI repositories concerning probability score of topics of the corresponding papers.}\label{tab:topic}
    \begin{tabular}{|c|l|>{\raggedright}p{0.38\textwidth}|r|r|}
    \hline
    \textbf{No.} & \textbf{Topic} & \textbf{Keywords} & \textbf{P-value} & \textbf{Cliff's $\delta$}\tabularnewline
    \hline
    \hline
    1 & Question Answering & text, question, word, visual, answer, classifier, query, language,
    attention, concept & $>$0.05 & 0.02 (N)\tabularnewline
    \hline
    2 & Transfer Learning & domain, target, description, source, particularly, classifier, create,
    transfer, tree, pair & $<$0.01 & -0.12 (N)\tabularnewline
    \hline
    3 & Representation Learning & learn, use, approach, method, task, network, show, base, representation,
    datum & $>$0.05 & -0.04 (N)\tabularnewline
    \hline
    4 & Event Tracking & descriptor, proposal, event, track, tracker, rank, window, million,
    outdoor, registration & $>$0.05 & -0.02 (N)\tabularnewline
    \hline
    5 & Image Processing & image, generate, shape, face, scene, pose, reconstruction, use, approach,
    patch & $>$0.05 & 0.01 (N)\tabularnewline
    \hline
    6 & Visual Recognition & embedding, subspace, discovery, crucial, operator, visual recognition,
    importantly, embed, basis, zero shot  & $>$0.05 & -0.01 (N)\tabularnewline
    \hline
    7 & Object Detection & object, video, dataset, region, segmentation, detection, action, motion,
    annotation, human & $>$0.05 & -0.05 (N)\tabularnewline
    \hline
    8 & Reinforcement Learning & agent, policy, version, constrain, tool, destribution, element, occur,
    privacy, theory & $<$0.01 & -0.11 (N)\tabularnewline
    \hline
    9 & Deep Network & model, network, propose, method, feature, deep, algorithm, show, image,
    base & $>$0.05 & -0.04 (N)\tabularnewline
    \hline
    10 & Object Correspondences & correspondence, flow, importance, group, object proposal, weakly supervise,
    tag, cluser, presence, treat & $>$0.05 & -0.04 (N)\tabularnewline
    \hline
    \end{tabular}
\end{table*}

Then, for each document, we leverage the topic model to compute a probability distribution of topics. Notice that each document is the abstract of an AI paper, which corresponds to an academic AI repository in our dataset. For each academic AI repository, we retrieve the corresponding document (i.e., the abstract of the corresponding paper). We represent the repository using the probability scores of the ten topics of the document. We view the ten scores as ten features of the repository. Finally, we apply the described method in RQ1 to analyze the relationship between the distribution of topics of an AI paper and the popularity of its accompanying academic AI repository. We apply the Wilcoxon rank-sum test and calculate Cliff's $\delta$ for each of the ten features. By doing so, we investigate whether popular and unpopular academic AI repositories are significantly different in terms of topics of the corresponding paper.

In Table~\ref{tab:topic}, we present p-values and Cliff's delta for comparing popular and unpopular academic AI repositories concerning the probability score of topics of the corresponding papers. As shown in the table, there are two topics with a p-value of less than 0.05. However, for both topics, the effect sizes are negligible. Hence, popular and unpopular academic AI repositories do not show significant difference with respect to topics of the corresponding papers. By comparing our experimental results in RQ1, we find that the software engineering practices of an academic AI repository have more impact on the popularity of the repository as compared to the topics of corresponding papers.

\vspace{0.2cm}\noindent \fbox{\parbox[c]{1\textwidth} {\emph{There is no significant association between the topic of an AI paper and the popularity of the accompanying academic AI repository. Topics of an AI paper are not likely to impact the popularity of its accompanying academic AI repository. The software engineering practices of an academic AI repository have more impact on popularity of the repository as compared to the topics of corresponding paper.}}}

\subsection{Images in the README file of academic AI repositories}\label{dis:image}

In our answer to RQ1, we find that popular and unpopular academic AI repositories show a significant difference with respect to the number of images shown in the README file. In our answer to RQ2, the experimental results show that the number of images in the README file is one of the most important features that distinguish popular academic AI repositories from unpopular ones. In this section, we further analyze the images that are shown in the README file of academic AI repositories. By doing so, we aim to provide more specific suggestions for AI researchers to present images in the README file of their academic AI repositories.

From the 1,040 labeled academic AI repositories, we find that 446 repositories show images in the README file. We perform a manual analysis of the images presented in the README file of the 446 repositories. We observe that academic AI repositories present two categories of images in the README file: (1) \emph{result images:} images that show running results of the code; (2) \emph{mechanism images:} images that describe the working mechanism of the code including architecture diagrams and math equations. For example, the README file of the  \texttt{vid2vid} repository\footnote{https://github.com/NVIDIA/vid2vid} presents seven animated images, which all illustrate how the proposed technique processes a video and output of the technique. These images belong to \emph{result images}. On the other hand, the README file of the \texttt{SENet} repository\footnote{https://github.com/hujie-frank/SENet} presents two images, which are architecture diagrams for describing the proposed deep networks. The two images belong to \emph{mechanism images}.

We count the number of academic AI repositories that present \emph{result images} and \emph{mechanism images} in the README file, respectively. We observe that 344 academic AI repositories present \emph{result images} in the README file, in which 132 repositories are popular. And 171 academic AI repositories present \emph{mechanism images} in the README file, in which 55 repositories are popular. We investigate whether there is a significant association between presenting \emph{result images} or \emph{mechanism images} in the README file and being a popular academic AI repository. Similarly to RQ1, for each image category, we calculate the number of popular and unpopular academic AI repositories that present and do not present images of the category, respectively. We apply Fisher's Exact test on the calculated numbers.

As a result, Fisher's Exact test shows that presenting \emph{result images} and presenting \emph{mechanism images} in the README file are both significantly associated with the popularity of academic AI repositories with a p-value of less than 0.001 and a p-value of less than 0.01, respectively. Both \emph{result images} and \emph{mechanism images} are encouraged to be presented in the README file of an academic AI repository.

\vspace{0.2cm}\noindent \fbox{\parbox[c]{1\textwidth} {\emph{Academic AI repositories present two categories of images in the README file, including images showing running results of the code and images describing the working mechanism of the code. Both categories of images are encouraged to be presented in the README file of an academic AI repository.}}}

\subsection{Implications for AI researchers}\label{dis:implication}

In Section~\ref{dis:topic}, we find that software engineering practices of an academic AI repository have more impact on popularity of the repository as compared to the topics of the corresponding papers. And in Section~\ref{sec:prepare}, we show that only a small proportion of academic AI repositories received many stars---suggesting that many academic AI repositories may not apply appropriate software engineering practices. According to our experimental results, we highlight the following \emph{software engineering practices} for AI researchers:

\vspace{0.1cm}\noindent\textbf{Provide code as detailed as possible and modularize the code.} Popular academic AI repositories provide code with more code files and code lines. Furthermore, popular academic AI repositories organize their code to more modules.

\vspace{0.1cm}\noindent\textbf{Pay special attention to the programming language and ML framework.} Academic AI repositories that use MATLAB are less likely to be popular. Furthermore, academic AI repositories that do not explicitly state the used framework in their README file are less likely to be popular. Academic AI repositories that use PyTorch, Torch, and MXNet are more likely to be popular. In addition, academic AI repositories that are owned by organizations and use Tensorflow are less likely to be popular.

\vspace{0.1cm}\noindent\textbf{Take more measures to ease reproducibility.} Popular academic AI repositories take more measures to ease the reproducibility of the paper results. They are more likely to provide pre-trained models and shell scripts for running their experiments. Furthermore, for organizational academic AI repositories, popular ones tend to provide datasets as compared to unpopular ones.

\vspace{0.1cm}\noindent\textbf{Improve the quality of documentation.} Popular academic AI repositories present better documentation. Their README files use more lists, code blocks, inline code elements to organize the textual contents. Their README files contain more images. Images that show running results of code and those describing working mechanisms of code are both encouraged to be presented in the README file of an academic AI repository. For organizational academic AI repositories, popular ones tend to provide more animated images than unpopular ones. Popular academic AI repositories also provide more links to other GitHub repositories in their README files. Their README files are more likely to include a URL to its project page and explicitly specify the used machine learning framework. Moreover, popular academic AI repositories are more likely to have a license.

\begin{figure}
  \centering
   \includegraphics[width=\textwidth]{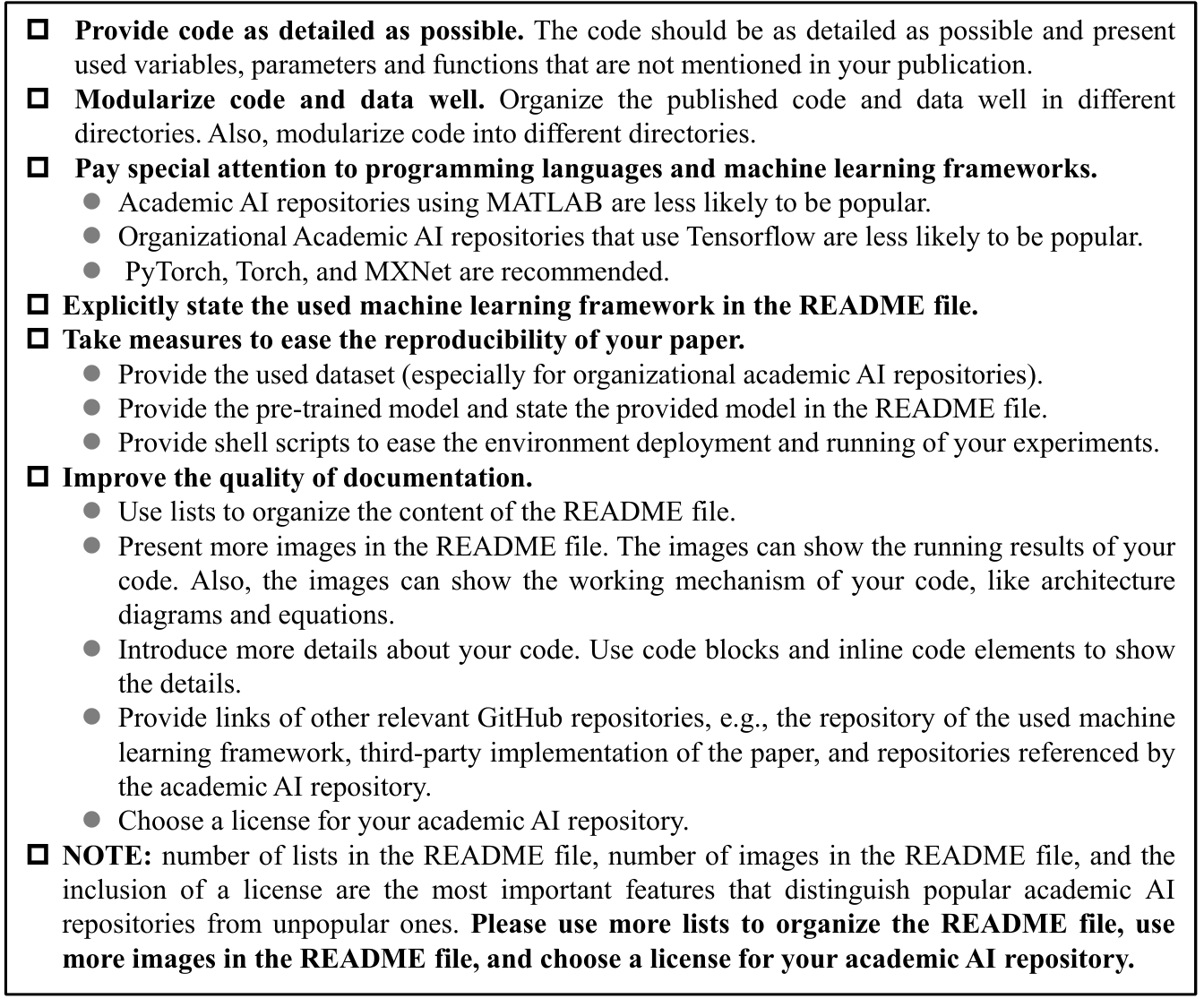}\\
  \caption{Checklist for AI researchers when publishing academic AI repositories.}\label{fig:checklist}
\end{figure}

Furthermore, the number of images in the README file, the number of links to other GitHub repositories in the README file, and the inclusion of a license are the most important features that are positively associated with the popularity of academic AI repositories. AI researchers should pay more attention to these documentation factors.

Based on our experimental results, we derive a checklist for AI researchers when publishing their academic AI repositories. Fig.~\ref{fig:checklist} presents the checklist.

\section{Threats to Validity}\label{discuss}

\vspace{0.1cm}\noindent\textbf{Threats to internal validity} are concerned with errors in our implementation code and dataset. We have double-checked the code. However, there may still exist some errors. The repositories in our dataset are selected from the \texttt{pwc} repository, which maintains a list of AI papers with references to their corresponding GitHub repositories. To make sure that the studied repositories are all academic AI repositories, we manually select the repositories that are released by authors of the corresponding papers. Hence, our dataset should be reliable. The calculations of the features such as \emph{has-model} and \emph{has-docker} are based on keyword matching. To validate whether our calculations of these features are correct, we manually checked 50 repositories that were randomly selected from our dataset. Our manual validation shows that our calculations of the features are correct.

Another threat is concerned with the labelling of the studied repositories. We label the top 20\% repositories that have the most number of stars as popular. The 20/80 boundary is commonly used for splitting skewed data in prior studies~\citep{phua2004minority,alves2010deriving,yan2017automating}. Furthermore, we use 10\% of the repositories as a gap between popular and unpopular academic AI repositories. By doing so, we make sure that popular and unpopular academic AI repositories have a considerable difference with respect to the number of stars: at least 55 stars.

Also, we focus on analyzing the documentation features of the README file in the root directory of academic AI repositories. Such repositories may have other documentation files. The extracted features from the additional documentation files may also be correlated with the popularity of academic AI repositories. Future studies are needed to investigate this possibility.

\vspace{0.1cm}\noindent\textbf{Threats to external validity} are concerned with the generality of our conclusions. The corresponding papers of the studied repositories are published in five different conferences, i.e., NeurIPS, ICML, CVPR, ICCV and ECCV. The five conferences are all ranked as A/A+ in the CORE rankings.\footnote{http://www.core.edu.au/conference-portal} Furthermore, our study is performed based on a dataset of 1,149 academic AI repositories, in which 1,040 (90\%) of the repositories are labeled as popular or unpopular. The remaining 109 (10\%) of the repositories are set as a gap between popular and unpopular repositories. With such a considerable number of academic AI repositories analyzed, we believe that our conclusion is convincing and reliable. Nevertheless, papers published in other AI conferences (e.g., IJCAI) and their accompanying academic AI repositories are not analyzed since the \texttt{pwc} repository only lists papers that are published in the five conferences. Future studies can verify our results using our experimental setup on more academic AI repositories whose corresponding papers are published in other conferences when such data becomes available.

Furthermore, we focus our analysis on academic AI repositories since we can retrieve a large number of such repositories thanks to the \texttt{pwc} repository.
Academic repositories in other areas are not analyzed due to lack of available data and it is unclear whether our results still hold. Most of our features can also be extracted from academic repositories in other areas. Hence, future studies can verify our results on academic repositories in other areas when data is available.

\vspace{0.1cm}\noindent\textbf{Threats to construct validity} are related to the suitability of the used evaluation measures. In RQ2, we use AUC to evaluate the effectiveness of the classification performance of the studied features. AUC is also a commonly used measure for evaluating the performance of classifiers~\citep{tian2015characteristics, fan2018chaff, fan2018early,han2019characterization}.

Furthermore, we calculate the \emph{has-model} feature by matching keywords such as ``trained model'' in the README file. We use the feature to characterize whether an academic AI repository provides pre-trained models. There may exist academic AI repositories that provide pre-trained models but not mention them in the README file. As described earlier, we manually analyzed 50 academic AI repositories that are randomly selected from our dataset. The analysis shows that the \emph{has-model} feature correctly reflects whether pre-trained models are provided in an academic AI repository. For academic AI repositories that mention the provided pre-trained models in the README file, they provide methods to retrieve the models. For academic AI repositories that do not mention pre-trained models in the README file, we cannot find a way to retrieve the pre-trained models that may be provided by the repositories. Thus, the calculation of the \emph{has-model} feature is not likely to threaten the validity of our results.

\section{Conclusion and Future Work}\label{conclusion}
In this paper, we analyze 1,149 academic AI repositories, aiming to suggest useful practices of popular academic AI repositories for AI researchers. We use 21 features to characterize the software engineering practices of the studied repositories. Then we compare the characteristics of popular academic AI repositories with unpopular ones. And we find that the two groups of academic AI repositories are statistically significantly different for 11 of the studied repository features---indicating that the two groups of repositories have significantly different software engineering practices. We also investigate the most important features that differentiate the two groups of repositories. According to our experimental results, we highlight good software engineering practices of popular academic AI repositories for AI researchers to improve their academic AI repositories

Our experimental results show that our features that characterize software engineering practices of academic AI repositories have more impact on the popularity of the repositories than topics of the corresponding papers. An automated tool can be built based on our features to predict whether a newly published academic AI repository will be popular. Such an automated tool could be useful for AI researchers. In addition, our experimental results show that popular academic AI repositories tend to take more measures to ease the reproducibility of results reported in the corresponding papers. It would be useful to investigate whether academic AI repositories can replicate the paper results and the potential difficulties of replicating the results. Furthermore, an AI paper can extend a prior paper. Thus, an academic AI repository may use code from another academic AI repository. It would also be useful to analyze how an academic AI repository leverages the code from other academic AI repositories. Our code and data are made publicly available, and researchers are encouraged to replicate and extend our work.

\begin{acknowledgement}
This research was partially supported by the National Key R\&D Program of China (No. 2018 YFB1003904) and the Australian Research Council's Discovery Early Career Researcher Award (DECRA) (DE200100021).
\end{acknowledgement}

\balance
\bibliographystyle{spbasic}
\bibliography{refs}

\begin{thebibliography}{57}
\providecommand{\natexlab}[1]{#1}
\providecommand{\url}[1]{{#1}}
\providecommand{\urlprefix}{URL }
\expandafter\ifx\csname urlstyle\endcsname\relax
  \providecommand{\doi}[1]{DOI~\discretionary{}{}{}#1}\else
  \providecommand{\doi}{DOI~\discretionary{}{}{}\begingroup
  \urlstyle{rm}\Url}\fi
\providecommand{\eprint}[2][]{\url{#2}}

\bibitem[{Aggarwal et~al.(2014)Aggarwal, Hindle, and Stroulia}]{aggarwal2014co}
Aggarwal K, Hindle A, Stroulia E (2014) Co-evolution of project documentation
  and popularity within github. In: Proceedings of the 11th Working Conference
  on Mining Software Repositories, ACM, pp 360--363

\bibitem[{Alves et~al.(2010)Alves, Ypma, and Visser}]{alves2010deriving}
Alves TL, Ypma C, Visser J (2010) Deriving metric thresholds from benchmark
  data. In: 2010 IEEE International Conference on Software Maintenance, IEEE,
  pp 1--10

\bibitem[{Balcan et~al.(2018)Balcan, Dick, Sandholm, and
  Vitercik}]{balcan2018learning}
Balcan MF, Dick T, Sandholm T, Vitercik E (2018) Learning to branch. In:
  International Conference on Machine Learning, pp 344--353

\bibitem[{Bissyand{\'e} et~al.(2013)Bissyand{\'e}, Thung, Lo, Jiang, and
  R{\'e}veillere}]{bissyande2013popularity}
Bissyand{\'e} TF, Thung F, Lo D, Jiang L, R{\'e}veillere L (2013) Popularity,
  interoperability, and impact of programming languages in 100,000 open source
  projects. In: 2013 IEEE 37th Annual Computer Software and Applications
  Conference, IEEE, pp 303--312

\bibitem[{Blei et~al.(2003)Blei, Ng, and Jordan}]{blei2003latent}
Blei DM, Ng AY, Jordan MI (2003) Latent dirichlet allocation. Journal of
  machine Learning research 3(Jan):993--1022

\bibitem[{Boettiger(2015)}]{boettiger2015introduction}
Boettiger C (2015) An introduction to docker for reproducible research. ACM
  SIGOPS Operating Systems Review 49(1):71--79

\bibitem[{Borges et~al.(2016{\natexlab{a}})Borges, Hora, and
  Valente}]{borges2016predicting}
Borges H, Hora A, Valente MT (2016{\natexlab{a}}) Predicting the popularity of
  github repositories. In: Proceedings of the The 12th International Conference
  on Predictive Models and Data Analytics in Software Engineering, ACM, p~9

\bibitem[{Borges et~al.(2016{\natexlab{b}})Borges, Hora, and
  Valente}]{borges2016understanding}
Borges H, Hora A, Valente MT (2016{\natexlab{b}}) Understanding the factors
  that impact the popularity of github repositories. In: 2016 IEEE
  International Conference on Software Maintenance and Evolution (ICSME), IEEE,
  pp 334--344

\bibitem[{Breiman(2001)}]{breiman2001random}
Breiman L (2001) Random forests. Machine learning 45(1):5--32

\bibitem[{Cliff(2014)}]{cliff2014ordinal}
Cliff N (2014) Ordinal methods for behavioral data analysis. Psychology Press

\bibitem[{Collberg and Proebsting(2016)}]{collberg2016repeatability}
Collberg C, Proebsting TA (2016) Repeatability in computer systems research.
  Communications of the ACM 59(3):62--69

\bibitem[{Collobert et~al.(2002)Collobert, Bengio, and
  Mari{\'e}thoz}]{collobert2002torch}
Collobert R, Bengio S, Mari{\'e}thoz J (2002) Torch: a modular machine learning
  software library. Tech. rep., Idiap

\bibitem[{Cutler et~al.(2012)Cutler, Cutler, and Stevens}]{cutler2012random}
Cutler A, Cutler DR, Stevens JR (2012) Random forests. In: Ensemble machine
  learning, Springer, pp 157--175

\bibitem[{Fan et~al.(2018{\natexlab{a}})Fan, Xia, Lo, and
  Hassan}]{fan2018chaff}
Fan Y, Xia X, Lo D, Hassan AE (2018{\natexlab{a}}) Chaff from the wheat:
  Characterizing and determining valid bug reports. IEEE Transactions on
  Software Engineering

\bibitem[{Fan et~al.(2018{\natexlab{b}})Fan, Xia, Lo, and Li}]{fan2018early}
Fan Y, Xia X, Lo D, Li S (2018{\natexlab{b}}) Early prediction of merged code
  changes to prioritize reviewing tasks. Empirical Software Engineering
  23(6):3346--3393

\bibitem[{Fan et~al.(2019)Fan, Xia, da~Costa, Lo, Hassan, and
  Li}]{fan2019impact}
Fan Y, Xia X, da~Costa DA, Lo D, Hassan AE, Li S (2019) The impact of changes
  mislabeled by szz on just-in-time defect prediction. IEEE Transactions on
  Software Engineering

\bibitem[{Fleiss(1971)}]{fleiss1971measuring}
Fleiss JL (1971) Measuring nominal scale agreement among many raters.
  Psychological bulletin 76(5):378

\bibitem[{Fogel(2005)}]{fogel2005producing}
Fogel K (2005) Producing open source software: How to run a successful free
  software project. " O'Reilly Media, Inc."

\bibitem[{Ghotra et~al.(2015)Ghotra, McIntosh, and
  Hassan}]{ghotra2015revisiting}
Ghotra B, McIntosh S, Hassan AE (2015) Revisiting the impact of classification
  techniques on the performance of defect prediction models. In: 2015 IEEE/ACM
  37th IEEE International Conference on Software Engineering, IEEE, vol~1, pp
  789--800

\bibitem[{Gousios et~al.(2014)Gousios, Pinzger, and
  Deursen}]{gousios2014exploratory}
Gousios G, Pinzger M, Deursen Av (2014) An exploratory study of the pull-based
  software development model. In: Proceedings of the 36th International
  Conference on Software Engineering, ACM, pp 345--355

\bibitem[{Gundersen et~al.(2017)Gundersen, Gil, and
  Aha}]{gundersen2017reproducible}
Gundersen OE, Gil Y, Aha DW (2017) On reproducible ai: Towards reproducible
  research, open science, and digital scholarship in ai publications. AI
  magazine 39(3):56--68

\bibitem[{Han et~al.(2019)Han, Deng, Xia, Wang, and
  Yin}]{han2019characterization}
Han J, Deng S, Xia X, Wang D, Yin J (2019) Characterization and prediction of
  popular projects on github. In: 2019 IEEE 43rd Annual Computer Software and
  Applications Conference (COMPSAC), IEEE, vol~1, pp 21--26

\bibitem[{Harrell~Jr(2015)}]{harrell2015regression}
Harrell~Jr FE (2015) Regression modeling strategies: with applications to
  linear models, logistic and ordinal regression, and survival analysis.
  Springer

\bibitem[{Hosmer~Jr et~al.(2013)Hosmer~Jr, Lemeshow, and
  Sturdivant}]{hosmer2013applied}
Hosmer~Jr DW, Lemeshow S, Sturdivant RX (2013) Applied logistic regression, vol
  398. John Wiley \& Sons

\bibitem[{Hu et~al.(2016)Hu, Zhang, Bai, Yu, and Yang}]{hu2016influence}
Hu Y, Zhang J, Bai X, Yu S, Yang Z (2016) Influence analysis of github
  repositories. SpringerPlus 5(1):1268

\bibitem[{Huang and Ling(2005)}]{huang2005using}
Huang J, Ling CX (2005) Using auc and accuracy in evaluating learning
  algorithms. IEEE Transactions on knowledge and Data Engineering
  17(3):299--310

\bibitem[{Jiang et~al.(2017)Jiang, Lo, He, Xia, Kochhar, and
  Zhang}]{jiang2017and}
Jiang J, Lo D, He J, Xia X, Kochhar PS, Zhang L (2017) Why and how developers
  fork what from whom in github. Empirical Software Engineering 22(1):547--578

\bibitem[{Kim et~al.(2004)Kim, Bergman, Lau, and Notkin}]{kim2004ethnographic}
Kim M, Bergman L, Lau T, Notkin D (2004) An ethnographic study of copy and
  paste programming practices in oopl. In: Proceedings. 2004 International
  Symposium on Empirical Software Engineering, 2004. ISESE'04., IEEE, pp 83--92

\bibitem[{Kimble(1992)}]{kimble1992plain}
Kimble J (1992) Plain english: A charter for clear writing. TM Cooley L Rev 9:1

\bibitem[{Li et~al.(2006)Li, Lu, Myagmar, and Zhou}]{li2006cp}
Li Z, Lu S, Myagmar S, Zhou Y (2006) Cp-miner: Finding copy-paste and related
  bugs in large-scale software code. IEEE Transactions on software Engineering
  32(3):176--192

\bibitem[{Newman et~al.(2010)Newman, Lau, Grieser, and
  Baldwin}]{newman2010automatic}
Newman D, Lau JH, Grieser K, Baldwin T (2010) Automatic evaluation of topic
  coherence. In: Human language technologies: The 2010 annual conference of the
  North American chapter of the association for computational linguistics, pp
  100--108

\bibitem[{Nosek et~al.(2015)Nosek, Alter, Banks, Borsboom, Bowman, Breckler,
  Buck, Chambers, Chin, Christensen et~al.}]{nosek2015promoting}
Nosek BA, Alter G, Banks GC, Borsboom D, Bowman SD, Breckler SJ, Buck S,
  Chambers CD, Chin G, Christensen G, et~al. (2015) Promoting an open research
  culture. Science 348(6242):1422--1425

\bibitem[{Paszke et~al.(2019)Paszke, Gross, Massa, Lerer, Bradbury, Chanan,
  Killeen, Lin, Gimelshein, Antiga et~al.}]{paszke2019pytorch}
Paszke A, Gross S, Massa F, Lerer A, Bradbury J, Chanan G, Killeen T, Lin Z,
  Gimelshein N, Antiga L, et~al. (2019) Pytorch: An imperative style,
  high-performance deep learning library. In: Advances in Neural Information
  Processing Systems, pp 8024--8035

\bibitem[{Phua et~al.(2004)Phua, Alahakoon, and Lee}]{phua2004minority}
Phua C, Alahakoon D, Lee V (2004) Minority report in fraud detection:
  classification of skewed data. Acm sigkdd explorations newsletter 6(1):50--59

\bibitem[{Portugal and do~Prado~Leite(2016)}]{portugal2016extracting}
Portugal RLQ, do~Prado~Leite JCS (2016) Extracting requirements patterns from
  software repositories. In: 2016 IEEE 24th International Requirements
  Engineering Conference Workshops (REW), IEEE, pp 304--307

\bibitem[{Prana et~al.(2018)Prana, Treude, Thung, Atapattu, and
  Lo}]{prana2018categorizing}
Prana GAA, Treude C, Thung F, Atapattu T, Lo D (2018) Categorizing the content
  of github readme files. Empirical Software Engineering pp 1--32

\bibitem[{Schober et~al.(2018)Schober, Boer, and
  Schwarte}]{schober2018correlation}
Schober P, Boer C, Schwarte LA (2018) Correlation coefficients: appropriate use
  and interpretation. Anesthesia \& Analgesia 126(5):1763--1768

\bibitem[{Scott and Knott(1974)}]{scott1974cluster}
Scott AJ, Knott M (1974) A cluster analysis method for grouping means in the
  analysis of variance. Biometrics pp 507--512

\bibitem[{Sonnenburg et~al.(2007)Sonnenburg, Braun, Ong, Bengio, Bottou,
  Holmes, LeCun, M{\~A}{\v{z}}ller, Pereira, Rasmussen
  et~al.}]{sonnenburg2007need}
Sonnenburg S, Braun ML, Ong CS, Bengio S, Bottou L, Holmes G, LeCun Y,
  M{\~A}{\v{z}}ller KR, Pereira F, Rasmussen CE, et~al. (2007) The need for
  open source software in machine learning. Journal of Machine Learning
  Research 8(Oct):2443--2466

\bibitem[{Sutton and Barto(2018)}]{sutton2018reinforcement}
Sutton RS, Barto AG (2018) Reinforcement learning: An introduction. MIT press

\bibitem[{Tantithamthavorn and Hassan(2018)}]{tantithamthavorn2018experience}
Tantithamthavorn C, Hassan AE (2018) An experience report on defect modelling
  in practice: Pitfalls and challenges. In: Proceedings of the 40th
  International Conference on Software Engineering: Software Engineering in
  Practice, ACM, pp 286--295

\bibitem[{Tantithamthavorn et~al.(2017)Tantithamthavorn, McIntosh, Hassan, and
  Matsumoto}]{tantithamthavorn2017empirical}
Tantithamthavorn C, McIntosh S, Hassan AE, Matsumoto K (2017) An empirical
  comparison of model validation techniques for defect prediction models. IEEE
  Transactions on Software Engineering 43(1):1--18

\bibitem[{Tantithamthavorn et~al.(2018)Tantithamthavorn, McIntosh, Hassan, and
  Matsumoto}]{tantithamthavorn2018impact}
Tantithamthavorn C, McIntosh S, Hassan AE, Matsumoto K (2018) The impact of
  automated parameter optimization on defect prediction models. IEEE
  Transactions on Software Engineering

\bibitem[{Tian et~al.(2015)Tian, Nagappan, Lo, and
  Hassan}]{tian2015characteristics}
Tian Y, Nagappan M, Lo D, Hassan AE (2015) What are the characteristics of
  high-rated apps? a case study on free android applications. In: 2015 IEEE
  International Conference on Software Maintenance and Evolution (ICSME), IEEE,
  pp 301--310

\bibitem[{Upton(1992)}]{upton1992fisher}
Upton GJ (1992) Fisher's exact test. Journal of the Royal Statistical Society:
  Series A (Statistics in Society) 155(3):395--402

\bibitem[{Wan et~al.(2017)Wan, Lo, Xia, Cai, and Li}]{wan2017mining}
Wan Z, Lo D, Xia X, Cai L, Li S (2017) Mining sandboxes for linux containers.
  In: 2017 IEEE International Conference on Software Testing, Verification and
  Validation (ICST), IEEE, pp 92--102

\bibitem[{Wan et~al.(2018)Wan, Xia, Hassan, Lo, Yin, and
  Yang}]{wan2018perceptions}
Wan Z, Xia X, Hassan AE, Lo D, Yin J, Yang X (2018) Perceptions, expectations,
  and challenges in defect prediction. IEEE Transactions on Software
  Engineering

\bibitem[{Wang et~al.(2018)Wang, Liu, Zhu, Liu, Tao, Kautz, and
  Catanzaro}]{NIPS2018_7391}
Wang TC, Liu MY, Zhu JY, Liu G, Tao A, Kautz J, Catanzaro B (2018)
  Video-to-video synthesis. In: Advances in Neural Information Processing
  Systems 31, pp 1144--1156

\bibitem[{Weber and Luo(2014)}]{weber2014makes}
Weber S, Luo J (2014) What makes an open source code popular on git hub? In:
  2014 IEEE International Conference on Data Mining Workshop, IEEE, pp 851--855

\bibitem[{Wilcoxon(1945)}]{wilcoxon1945individual}
Wilcoxon F (1945) Individual comparisons by ranking methods. Biometrics
  bulletin 1(6):80--83

\bibitem[{Woodfield et~al.(1981)Woodfield, Dunsmore, and
  Shen}]{woodfield1981effect}
Woodfield SN, Dunsmore HE, Shen VY (1981) The effect of modularization and
  comments on program comprehension. In: Proceedings of the 5th international
  conference on Software engineering, IEEE Press, pp 215--223

\bibitem[{Xia et~al.(2019)Xia, Wan, Kochhar, and Lo}]{xia2019practitioners}
Xia X, Wan Z, Kochhar PS, Lo D (2019) How practitioners perceive coding
  proficiency. In: 2019 IEEE/ACM 41st International Conference on Software
  Engineering (ICSE), IEEE, pp 924--935

\bibitem[{Yan et~al.(2017)Yan, Xia, Zhang, Yang, and Xu}]{yan2017automating}
Yan M, Xia X, Zhang X, Yang D, Xu L (2017) Automating aggregation for software
  quality modeling. In: 2017 IEEE International Conference on Software
  Maintenance and Evolution (ICSME), IEEE, pp 529--533

\bibitem[{Yan et~al.(2018)Yan, Xia, Shihab, Lo, Yin, and
  Yang}]{yan2018automating}
Yan M, Xia X, Shihab E, Lo D, Yin J, Yang X (2018) Automating change-level
  self-admitted technical debt determination. IEEE Transactions on Software
  Engineering

\bibitem[{Yang et~al.(2018)Yang, Lu, Lee, Batra, and Parikh}]{yang2018graph}
Yang J, Lu J, Lee S, Batra D, Parikh D (2018) Graph r-cnn for scene graph
  generation. In: Proceedings of the European Conference on Computer Vision
  (ECCV), pp 670--685

\bibitem[{Zar(2005)}]{zar2005spearman}
Zar JH (2005) Spearman rank correlation. Encyclopedia of biostatistics 7

\bibitem[{Zhu et~al.(2014)Zhu, Zhou, and Mockus}]{zhu2014patterns}
Zhu J, Zhou M, Mockus A (2014) Patterns of folder use and project popularity: A
  case study of github repositories. In: Proceedings of the 8th ACM/IEEE
  International Symposium on Empirical Software Engineering and Measurement,
  ACM, p~30

\end{thebibliography}

\end{document}